\documentclass[twocolumn]{openjournal}

\usepackage{xcolor}
\usepackage{textgreek}
\usepackage[utf8]{inputenc}
\usepackage[english]{babel}
\usepackage{csquotes}

\usepackage{hyperref}
\hypersetup{
    unicode, 
    colorlinks=true,
    linkcolor=linkcolor,
    citecolor=linkcolor,
    filecolor=linkcolor,
    urlcolor=linkcolor,
}
\usepackage{color,colortbl}
\definecolor{linkcolor}{rgb}{0.0,0.3,0.5}
\DeclareGraphicsExtensions{.bmp,.png,.jpg,.pdf}
\usepackage{verbatim}
\usepackage[normalem]{ulem}
\usepackage{orcidlink}
\usepackage{soul}
\usepackage{enumitem} % 022426
\usepackage{amsmath}  % 030926
\usepackage{hyperref} % 031026
\usepackage{cleveref}
\usepackage{import}   % 040926
\graphicspath{ {./figs/} } % def path to figs
\usepackage{longtable}
\usepackage{bm}
\usepackage{float}
\usepackage{xcolor}
\usepackage{soul}

\usepackage{tabularx} % Required for paragraph-style columns with automatic width

\usepackage{needspace} % 110726

\begin{document}

%%%   %%%   %%%
%%%   %%%   %%%
%%%   %%%   %%%
%% Title Page Information
% Title
%\title{The HOD Floor and Ceiling of small-scale clustering constraints}
\title{Bounding the Effect of HOD Assumptions on Small-Scale Clustering Constraints}

% Primary Author - Nick Magnelli
\author{Nick Magnelli \orcidlink{0000-0001-5555-1182}}
\email{magnelli@ksu.edu}
\affiliation{Department of Physics, Kansas State University, Manhattan, KS}

% Secondary Author - Zachery Brown 
\author{Zachery Brown \orcidlink{0000-0001-8208-7282
}}
\affiliation{Department of Physics, Kansas State University, Manhattan, KS}

% Secondary Author - Lado Samushia
\author{Lado Samushia \orcidlink{0000-0002-1609-5687
}}
\affiliation{Department of Physics, Kansas State University, Manhattan, KS}
\affiliation{Abastumani Astrophysical Observatory, Tbilisi, GE-0179, Georgia}

%%%   %%%   %%%
%%%   %%%   %%%
%%%   %%%   %%%
%% Abstract, Keywords, Make Title
% Abstract
\begin{abstract}
\noindent
Small-scale galaxy clustering is expected to contain substantial cosmological information, but the extent to which this information constrains halo-based cosmologies independent of an assumed galaxy--halo connection remains unclear. The primary goal of this information-content study is to quantify the difference in clustering constraints that result from assuming the non-cosmological contribution is absolutely minimized versus maximized given some otherwise-fixed methodology. To this end we compare two limiting treatments, the \enquote{floor} and \enquote{ceiling}, of the standard five-parameter HOD model by analyzing two-point correlation function multipoles on scales ranging from $5$--$80$ Mpc/$h$ in LRG-like mock galaxy catalogs built from 81 cosmologies in the \textsc{AbacusSummit} suite. In the conservative floor case, we impose only broad initial bounds and profile over HOD parameters to determine the minimum constraining power available given the assumed HOD model; we accomplish this with {\tt HODmin}, a two-stage global optimization algorithm written for minimizing $\chi^2$ in HOD space. In the upper-limiting ceiling case, we assume the HOD parameters are known exactly. We generally find a significant difference between the floor and ceiling when comparing against the same Planck $\Lambda$CDM mock data vector with identical modeling assumptions. Many cosmologies agree well with data in the floor, and yet in the ceiling are excluded by multiple orders of magnitude in $\chi_\nu^2$. We therefore observe the strength of small-scale clustering constraints depends heavily on the amount of prior HOD information assumed, even above the regime where the one-halo term dominates. Our generally wide floor--ceiling bracket suggests strong small-scale clustering constraints may arise largely due to non-cosmological priors, and motivates a detailed follow-up study on DESI data.
\end{abstract}

% Keywords

% Make Title
\maketitle

%%%   %%%   %%%
%%%   %%%   %%%
%%%   %%%   %%%
% Section: Introduction
%%%   %%%   %%%
%%%   %%%   %%%
%%%   %%%   %%%
% Introduction
\section{Introduction}\label{sec:intro}

The study of galaxy clustering seeks to extract cosmological information from the spatial distribution of galaxies, and the spectroscopic galaxy surveys driving the field have grown considerably in the past two decades. The Dark Energy Spectroscopic Instrument (DESI), in its second data release (DESI--DR2), compiled over 30 million galaxies in the largest public spectroscopic galaxy catalog to date \citep{DESI_DR2_II}. The third DESI data release (DESI--DR3) will be larger still \citep{desi_sv_validation_2024}, and several upcoming or proposed spectroscopic surveys are expected to further sharpen constraining power, including MegaMapper \citep{schlegel2022megamapper}, the MUltiplexed Survey Telescope \citep{zhao2024multiplexed}, the Maunakea Spectroscopic Explorer \citep{sheinis2023maunakea}, the Wide-Field Spectroscopic Telescope \citep{dassignies_zhao_yu_kneib_2023,bacon2024wstwidefieldspectroscopic}, Roman \citep{wang2022high,eifler2024cosmology}, Euclid \citep{laureijs2011euclid,tutusaus2020euclid,mellier2024euclid}, and Spec-S5 \citep{besuner2025spectroscopic}.

The most robust constraints from galaxy clustering on cosmological models and parameters currently tend to come from large-scale features in the two-point correlation function (2PCF), especially the baryon acoustic oscillation (BAO) signature and redshift-space distortions (RSDs). The BAO signature is a statistical excess of galaxy pairs separated by a particular redshift-dependent distance scale \citep{eisenstein_2005,DESI_DR2_II}. RSDs are the apparent, model-dependent anisotropies in galaxy clustering between the line-of-sight and transverse directions \citep{kaiser1987clustering,hamilton1998linear,scoccimarro2004redshift,percival2009testing,samushia2012interpreting}. DESI-based BAO and RSD analyses considerably tightened the cosmological constraints derived from the Sloan Digital Sky Survey (SDSS), especially when combined with cosmic microwave background (CMB) data \citep{eisenstein_2005,desi_2024_vi_bao_2025,DESI_DR2_II}.

While the large-scale BAO and RSD analyses are generally considered the most reliable clustering probes to test models, substantial cosmological information is often reported to exist on smaller scales as well (e.g. \citealt{wang_2017_small_scales_mnras,2022MNRAS.510.5376S, yuan_garrison_eisenstein_wechsler_2022, desi_dr1_fullshape_2024}). A number of studies have sought to extract this information.
The notable recent work from the DESI collaboration \citep{desi_dr1_fullshape_2024} examined the first data release (DR1) full-shape clustering of galaxies, quasars, and Lyman-$\alpha$ forest tracers analytically down to about 30 Mpc/$h$ and concluded by tightening the already-strong cosmological parameter constraints deduced from the DESI DR1 BAO data by a nontrivial margin. However, computing observables that accurately represent the predictions of expanding cosmologies is far more complicated at small scales than at large scales. At large scales, analytical predictions computed with cosmological perturbation theory (PT) are widely accepted in the literature for models based on the Friedmann--Lema$\mathrm{\hat{\i}}$tre--Robertson--Walker (FLRW) metric, for such models generally assume the large-scale density field lacks any substantial coupling to the non-perturbative density variations present on small scales (e.g. \citealt{ivanov2024effective}).\footnote{There is, however, a growing body of literature questioning whether cosmological backreaction complicates the widely-accepted notion of Friedmannian expansion, which would consequently affect the application of PT thereto. Several studies have argued that backreaction has no significant effect: \citet{ishibashi2006can,green2011new,green2013examples,green2014well}. Many studies have argued to the contrary: \citet{buchert2000average,buchert2000back,rasanen2004dark,rasanen2006accelerated,kolb2006cosmic,wiltshire2007exact,buchert2008dark,clarkson2011does,buchert2012backreaction,buchert2015there}. For reviews, see \citet{bolejko2017inhomogeneous,schander2021backreaction}.} At the sufficiently small scales of interest, the requisite perturbative assumptions break down \citep{yuan_garrison_eisenstein_wechsler_2022,lange_etal_2022,baleato_lizancos_seljak_karamanis_2025}. Consequently, we must engage non-perturbative means to compute model predictions. This generically requires adding assumptions to the model being tested, which in turn often involves introducing nuisance parameters; indeed, the aforementioned DESI-DR1 full-shape study invoked 42 nuisance parameters in their methodology \citep{desi_dr1_fullshape_2024}. Moreover, in small-scale clustering, the FLRW-based physics cannot be cleanly separated from the non-FLRW nuisance framework, nor can model predictions be separated from the initial conditions upon which any of these physical or nuisance assumptions evolve, insofar as different initial conditions lead to numerically distinct predictions (cf. Section~\ref{sec:results:ph002} and Section~\ref{sec:results:phase24}). Thus, a \enquote{cosmological model} that we test in small-scale clustering is the synthesis of these assumptions regarding the FLRW physics, the non-FLRW nuisance framework, and the initial conditions.\footnote{We will address this matter thoroughly in future work. In this work, we will follow the precedent of the literature in applying the terms \enquote{model} and \enquote{cosmology} in a loose, context-evident manner, except when ascribing constraints, in which case we will be explicit.} 

Many methods have been proposed and employed to compute model predictions for small-scale clustering observables. One such method is the use of high-fidelity numerical simulations to produce the cold dark matter (CDM) halo distribution predicted by CDM-based models, supplemented with an assumed nuisance model that maps the halo distribution onto a galaxy distribution. Numerous studies have employed such methodologies to deduce cosmological constraints. \citet{yuan_garrison_eisenstein_wechsler_2022} derived \enquote{stringent $\sigma_8$ constraints} from BOSS CMASS DR12 LRGs over the range 0 to 30 Mpc/$h$ in an {\tt \textsc{AbacusSummit}}-based hybrid MCMC+emulator methodology which accounted for multiple galaxy--halo connection extension models. 
\citet{lange_etal_2022} also concluded with strong constraints on $\sigma_8$ while considering similar nuisance extensions, but by modeling with the {\tt \textsc{Aemulus}} simulation suite and testing against BOSS LOWZ data over 0.4 to 63 Mpc/$h$.
\citet{2025arXiv251215962L} achieved strong constraints through an {\tt \textsc{AbacusSummit}}-based template with complex nuisance models when testing against DESI-DR1 LRGs down to scales of 0.4 Mpc/h, noting their results were in agreement with other recent studies (see references in \enquote{Section 5} thereof).
This general approach, however, is likewise not without noteworthy complications. High-fidelity simulations are sufficiently expensive that even the largest suites with substantial per-simulation volume, such as {\tt \textsc{AbacusSummit}} and {\tt \textsc{Aemulus}}, can only sample the cosmological parameter space sparsely relative to the number of cosmological parameters \citep{maksimova2021abacussummit,derose2019aemulus}.
The allowed galaxy--halo mapping flexibility scales with the addition of free nuisance parameters, such that a single CDM-only simulation can lead to an arbitrarily wide range of clustering predictions according to the nuisance preferences of the modeler.
Direct computation of observables in this manner is costly, which complicates an attempt at exhaustive inference. Emulation may be employed to improve efficiency, but the finite training set introduces interpolation error. The pseudorandom nature of galaxy population, the irregular dependence of observables on nuisance parameters, and practical limits on training density can all contribute to non-smooth likelihood surfaces with many local minima.

A distinct but similarly important consideration regards how nuisance parameters are treated statistically. The most common approach employed in the literature is to marginalize over nuisance parameters and compute a posterior distribution for the cosmological parameters (e.g. for PT-based analyses, see \citealt{white5435tests,2021PhRvD.103d3525C,desi_dr1_fullshape_2024,desi_fullshape_2025,carrilho2023cosmology,simon2023consistency,holm2023bayesian}; for the effect of Jeffreys priors, see \citealt{donald2023analysis,hadzhiyska2023cosmology,gsponer2024cosmological}; for the effect of HOD-informed priors, see \citealt{akitsu2024mapping,ivanov2024full,zhang2025hod}; for simulation-based applications, see \citealt{lange_etal_2022,yuan_garrison_eisenstein_wechsler_2022,2026ApJ...998..268G}). An alternative approach is to profile nuisance parameters: for each cosmological model, one finds the nuisance parametrization yielding the best fit to data, and the best-fit cosmological models are then compared to one another (e.g. \citealt{holm2023bayesian,2024MNRAS.535.3686B,morawetz2025frequentist,2025PhRvD.111h3504H}). Marginalization and profiling are intended to address different questions. Marginalization considers what effect data produces on the prior parameter space of a model, whereas profiling considers whether any nuisance parametrization exists to provide a good fit between model and data. This distinction is especially important in small-scale clustering. Variations in the galaxy--halo connection can induce 2PCF changes exceeding those caused by typical variations in cosmological parameters, and with the aforesaid difficulties in modeling the galaxy--halo connection, this can induce strong and undesired prior effects. A conservative alternative approach is to address whether specific cosmological models can be ruled out even when the nuisance parameters are allowed maximal variability.

The primary goal of this paper is, given some methodology to derive cosmology constraints from small-scale galaxy clustering, to quantify the difference between assuming the contribution from non-cosmological information is absolutely minimized versus absolutely maximized. This primary goal requires at least three intermediate goals based on the chosen methodology. First, we must derive constraints by assuming the contribution from non-cosmological information is minimized. Second, we must derive constraints by assuming the contribution from non-cosmological information is maximized. Third, we must compare the results from the first and second intermediate goals. To attain the first intermediate goal we require high-fidelity numerical simulations, an assumed galaxy--halo connection model upon which our results are predicated, and a profile likelihood over nuisance parameters. We proceed to explain why each of these three elements is required, as well as our subsequent implementation.

Deriving constraints that minimize the contribution from non-cosmological information requires deriving constraints that maximize cosmological information, which in turn requires maximizing the accuracy of model prediction computations. Since accurate model predictions cannot be achieved for small-scale galaxy clustering through analytical means, we must employ non-analytical means. This establishes a give-and-take between the accuracy and completeness of the prediction, since the inability to complete a prediction analytically renders the model prediction computations incomplete until the model is supplemented with additional assumptions that are non-cosmological by definition. Since we are maximizing accuracy at the expense of completeness, we must employ the most accurate class of computations, which are high-fidelity numerical CDM-only simulations based on entirely collisionless fluid equations. Within this class we use the {\tt \textsc{AbacusSummit}} simulations.

Although considered accurate for the CDM distribution of a given cosmology, {\tt \textsc{AbacusSummit}} alone cannot predict the observable galaxy distribution. To this end we must assume an HOD model that takes the DM-distribution and returns a galaxy distribution. Moreover, since it remains unknown whether any particular HOD model follows as a necessary consequence of any particular cosmological model, our constraints are predicated upon the assumed HOD model.

By definition, minimizing non-cosmological information requires a profile likelihood to locate the HOD parametrization that globally minimizes the statistical tension between model and data. This determines the minimum amount of constraining power available on the cosmology given the assumed HOD model and methodology. Profiling, not marginalization, must be employed to address our primary goal.
 
For each {\tt \textsc{AbacusSummit}}-based cosmological model, we seek the galaxy--halo nuisance parametrization, specifically of the standard five-parameter halo occupation distribution (HOD) model, that globally minimizes $\chi^2$ in comparison to mock data. This yields globally minimized goodness-of-fit constraints on cosmological models, where such models are considered viable if there exists any allowed HOD parametrization that fits well to data. Such constraints, moreover, are independent of any prior likelihood surface.

Moreover, our first intermediate goal, and hence our primary goal, requires at least a profile likelihood over HOD parameters within high-fidelity numerical simulations. Existing studies which lack this ingredient are approaching different goals altogether. We have found very few studies in literature review which attempt even a similar methodology.
\citet{alam2020multitracer} profiled nuisance parameters in the MultiDark Planck simulation to determine the nuisance model yielding the best fit to eBOSS data. \citet{yuan2021evidence} employed a technique called \enquote{covariance matrix adaptation evolution strategy} \citep{hansen2016cma} for global optimization in a simulation-based profile likelihood analysis, testing eight {\tt \textsc{AbacusCosmos}} simulations and five {\tt \textsc{AbacusSummit}} simulations against BOSS--CMASS data. \citet{yuan2022abacushod}, in their introduction of the {\tt \textsc{AbacusHOD}} algorithm, applied the same methodology to eBOSS data for one simulation. \citet{yuan2024desi} applied a similar approach in testing for galaxy assembly bias and satellite radial profile bias, likewise for a single simulation. Our work continues this program of simulation-based profile likelihood analysis by extending to most of the {\tt \textsc{AbacusSummit}} step and emulator grids,
studying the dependence of our results on phase, HOD, scale cuts, and multipole choice, and, most importantly, validating the convergence of our global optimization procedure explicitly.
This latter point is not trivial, insofar as cosmology-maximizing constraints require an accurate global minimization. We have found no mention in the simulation + profile papers listed above of a methodology to validate the global minimization algorithm employed, such that even these cannot be said to attempt addressing our primary goal. This apparent lack of relevant papers can likely be attributed to the relatively recent availability of high-fidelity, large-volume CDM-only simulations coupled with the noticeable minority of minimization-based statistical studies compared to marginalization-based Bayesian statistics.

To carry out the required minimization, we introduce {\tt HODmin}, a two-stage global optimization algorithm that combines adaptive pseudorandom domain compression with local optimization. The first stage searches broadly through the HOD parameter space while progressively narrowing the allowed region, and the second stage refines the result with a local minimizer. This flexible and tunable algorithm is designed for irregular, non-smooth $\chi^2$ surfaces of the kind produced by simulation-based HOD modeling. 

We apply {\tt HODmin} to mock luminous red galaxy (LRG) catalogs built from the {\tt \textsc{AbacusSummit}} suite. LRGs are a natural target because of their importance in current spectroscopic surveys like DESI and their good completeness \citep{zhou2023desi}. We are especially interested in two limiting cases. In the conservative case, or the \enquote{floor}, we impose only broad initial HOD bounds, along with modest filters on number density and satellite fraction, to compare cosmological models through the aforementioned method of profiling HOD parameters that yields globally minimized $\chi^2$ values. In the upper-limiting case, or the \enquote{ceiling}, we assume the HOD parametrization is known exactly and compute the goodness-of-fit $\chi^2$ without any profiling. 
The floor and ceiling impose bounds on the range of constraining power available, conditional upon the adopted five-parameter HOD model (Section~\ref{sec:prereq:HOD}) and methodological framework (Section~\ref{sec:Methods}).
As we show in Section~\ref{sec:results:floor_ceiling}, this bracket is wide. With $8~(\mathrm{Gpc}/h)^3$ of data generated from the Planck best-fitting $\Lambda$CDM model (hereafter \enquote{Planck $\Lambda$CDM}; see \citealt{aghanim2020planck}), the default phase, a fiducial galaxy--halo connection, and a common set of modeling assumptions (Section~\ref{sec:Methods:Procedure}), we find that $\sim81\%$ of the {\tt \textsc{AbacusSummit}} simulations are excluded at $3\sigma$ in the upper-limiting ceiling case, compared to only $25\%$ in the conservative floor case (Section~\ref{sec:results:floor_ceiling}). Many of these cosmologies agree substantially well with data when the HOD model is given sufficient freedom to vary, and yet are excluded by multiple orders of magnitude in $\chi^2$ when the HOD model is fixed to the assumed parametrization. Thus, while small-scale clustering contains potentially substantial cosmological information, the extent to which it can be robustly extracted depends strongly on the assumptions made regarding nuisance parameters. The floor case also exposes a substantial degeneracy in HOD space which indicates the HOD model cannot be identified from its 2PCF multipoles alone, even when the cosmology is known otherwise (Figure~\ref{fig:corner_val}). 

This paper is organized as follows. In Section~\ref{sec:prereq} we introduce the simulations, observables, HOD model, covariance matrices, and statistical framework used throughout our work. In Section~\ref{sec:HODmin} we describe the {\tt HODmin} algorithm. In Section~\ref{sec:Methods} we describe the mock analysis, validation pipeline, and tests. We present our results in Section~\ref{sec:results}, and discuss their implications in Section~\ref{sec:discussion}, where we conclude.

%%%   %%%   %%%
%%%   %%%   %%%
%%%   %%%   %%%
% Section: Prerequisites
%%%   %%%   %%%
%%%   %%%   %%%
%%%   %%%   %%%
% Prerequisites

\section{Prerequisites}
\label{sec:prereq}

We begin by introducing the simulations, observables, HOD model, covariance matrices, and statistical framework used throughout the paper.

%%%   %%%   %%%
%%%   %%%   %%%
%%%   %%%   %%%
% Prerequisites -- AbacusSummit
\subsection{Simulations}
\label{sec:prereq:sims}

Our analysis is based on the {\tt \textsc{AbacusSummit}} suite of high-resolution cosmological $N$-body simulations. {\tt \textsc{AbacusSummit}} provides both large-volume simulations, used here to model clustering in different cosmologies, and smaller-volume simulations, used here to estimate covariance matrices. The suite includes 150 \enquote{cubic boxes} of side length $2 \: \mathrm{Gpc}/h$ spanning 97 unique parametrizations of the $w_0w_a$CDM cosmology, as well as 1883 \enquote{small boxes} of side length $500 \: \mathrm{Mpc}/h$ corresponding to different initial conditions, or \enquote{phases}, of the Planck $\Lambda$CDM model \citep{maksimova2021abacussummit}. We use only 81 of the 97 unique $w_0w_a$CDM parametrizations due to various technical issues that arose in pre-processing.\footnote{These technical issues, such as files not existing and permission errors, were neither expedient nor necessary to address for the sake of this proof-of-concept analysis.} These 81 parametrizations are listed in Table~\ref{tab:HODmin_fin}. We use the standard {\tt \textsc{AbacusSummit}} notation throughout this work: cosmologies are labeled by \enquote{{\tt c}} followed by a three-digit code, and phases by \enquote{{\tt ph}} followed by a three-digit code for cubic boxes or a four-digit code for small boxes. For example, \enquote{{\tt c000\_ph000}} denotes the default phase of the Planck $\Lambda$CDM model.\footnote{A complete list of {\tt \textsc{AbacusSummit}} simulations is given at \url{https://abacussummit.readthedocs.io/en/latest/simulations.html}.}

The eight $w_0w_a$CDM parameters are $\omega_b$, $\omega_\mathrm{cdm}$, $\sigma_8$, $n_s$, $\alpha_s$, $N_\mathrm{eff}$, $w_0$, and $w_a$. $\omega_b$ and $\omega_\mathrm{cdm}$ denote the physical baryon and CDM densities, $\sigma_8$ is the amplitude of structure, $n_s$ is the scalar spectral tilt, $\alpha_s$ is the running of $n_s$, and $N_\mathrm{eff}$ is the effective number of relativistic species. The parameters $w_0$ and $w_a$ describe the dark energy equation of state through
\begin{equation}
    w(a) = w_0 + (1-a)w_a ,
\end{equation}
where $a$ is the scale factor. The standard $\Lambda$CDM case corresponds to $w_0=-1$ and $w_a=0$; for more complete discussions of $w_0w_a$CDM and similar parametrization, see: \citet{chevallier2001accelerating,linder2003exploring,caldwell2005limits,albrecht2006report,wang2008figure,scherrer2015mapping,lee2026comparing}.

The {\tt \textsc{AbacusSummit}} cosmologies include the Planck $\Lambda$CDM model ({\tt c000}) and other special parametrizations ({\tt c[001-004,009-022]}), a step grid ({\tt c[100-126]}) in which cosmological parameters are varied approximately one at a time around the Planck $\Lambda$CDM cosmology, and an emulator grid ({\tt c[130-181]}) in which parameters are sampled over an 8-dimensional ellipsoid \citep{maksimova2021abacussummit}. The step-grid simulations are often useful for derivative estimates. In this work, however, we do not use them for derivative calculations; instead, we treat all simulations herein as unique models for which goodness-of-fit can be computed.

%%%   %%%   %%%
%%%   %%%   %%%
%%%   %%%   %%%
% Prerequisites -- HOD

\subsection{Galaxy--halo connection}
\label{sec:prereq:HOD}

The {\tt \textsc{AbacusSummit}} simulations are designed to predict the clustering of halos for each corresponding cosmological model. However, the observed tracers are galaxies, not halos, and therefore we require a prescription to connect the simulated halo distribution to a galaxy distribution. We accomplish this with an HOD model, which assigns galaxies to halos according to some parametric framework \citep{2002ApJ...575..587B,2003ApJ...593....1B,2005ApJ...633..791Z,2009ApJ...707..554Z,2018ARA&A..56..435W}. These HOD parameters are the nuisance parameters over which we profile in our analysis.

We use the standard five-parameter HOD model for LRGs, often called the \enquote{vanilla} HOD model. In this model, galaxies are classified as centrals or satellites. The probability that a halo hosts a central galaxy is
\begin{equation}
\label{equ:Ncen}
    N_{\mathrm{cen}} = \frac{1}{2} \left[ 1 + \mathrm{erf} \left( \frac{\mathrm{log}M-\mathrm{log}M_{\mathrm{cut}}}{\sqrt{2}\sigma} \right)  \right] \,,
\end{equation}
where $\mathrm{erf}$ is the error function, $M$ is the halo mass, and $\mathrm{log}M_{\mathrm{cut}}$ and $\sigma$ are free HOD parameters. The characteristic number of satellites hosted by a halo is
\begin{equation}
\label{equ:Nsat}
    N_{\mathrm{sat}} = N_{\mathrm{cen}} \left( \frac{M-\kappa M_{\mathrm{cut}}}{M_1} \right)^{\alpha}  \,,
\end{equation} 
where $\mathrm{log}M_1$, $\alpha$, and $\kappa$ are also free HOD parameters \citep{ yuan_garrison_eisenstein_wechsler_2022,yuan2022abacushod}. We refer to this five-parameter model as the \enquote{5pHOD} model, and any set of five particular 5pHOD parameters as a \enquote{5pHOD-set}. Each 5pHOD-set applied to an {\tt \textsc{AbacusSummit}} simulation implies both a galaxy number density, $\bar{n}$, and a satellite fraction, $f_\mathrm{sat}$. In principle, $\bar{n}$ is directly observable and could be included as an additional data point in the fit. We omit this choice herein, and instead impose broad cuts on $\bar{n}$ and $f_\mathrm{sat}$ in various places to remove the contribution of extreme catalogs that are not representative of LRG-like samples (Section~\ref{sec:HODmin}, Section~\ref{sec:Methods:catalogs_interp}, Section~\ref{sec:Methods:Repo:Mocks}). 

One might object to our using the 5pHOD model on the grounds of exceeding simplicity. We do not deny the complexities of HOD modeling and the value of extensions, but for various reasons, both practical and theoretical, our methodology does not require us to employ more complicated HOD models. On the practical side, we expedite the derivation of cosmology-maximizing constraints when we assume a less complicated HOD model with a weaker effect on our observables. Also, global optimization convergence can be verified more reliably on a simpler HOD model with fewer nuisance parameters. On the theoretical side, our work focuses on information content given some HOD model, rather than improving HOD modeling or small-scale clustering constraints. By construction, our results are deliberately conditional upon our HOD modeling choices. Even to the HOD modeler who favors extended HOD models, an information-content study on the 5pHOD model is not an irrelevant undertaking. 
By analyzing the 5pHOD model separately, we isolate the information contained therein from the information contained within entire 5pHOD-based extension models, some of which are not nested within one another but are merely overlapping, such as velocity-bias-only and assembly-bias-only extensions.
Hence, in principle, our results will bear upon any arbitrarily extended model that contains 5pHOD as a nested subset, for the information contained within 5pHOD will be contained within such an extension. On the contrary, an information-theory study that considers only a highly-extended HOD model would not apply straightforwardly to the less complicated models nested therewithin, let alone to merely overlapping models with altogether distinct degrees of freedom.
Moreover, if our results indicate that intermediate-to-small-scale clustering constraints depend heavily on the \enquote{vanilla} 5pHOD assumptions, then one might reasonably expect the extension assumptions to contribute nontrivially as well. At the very least, such an expectation can then be tested separately for any particular 5pHOD-based extension model; it is beyond the purpose of this proof-of-concept mock study to test separately the many different 5pHOD-based extensions employed in the literature. We are content with testing the 5pHOD core common to all 5pHOD-based extensions and saving a more detailed study of such extensions for a follow-up data analysis.

We implement the 5pHOD model using {\tt \textsc{AbacusHOD}}, an algorithm which efficiently populates {\tt \textsc{AbacusSummit}} halos with galaxies given a 5pHOD-set. {\tt \textsc{AbacusHOD}} populates galaxies through pseudorandom number generation, and therefore requires that a seed be fixed. {\tt \textsc{AbacusHOD}} also makes several efficiency choices that differ from the exact 5pHOD model \citep{yuan2022abacushod}. Thus, to distinguish from the exact 5pHOD model, we refer hereafter to the {\tt \textsc{AbacusHOD}} implementation of the 5pHOD model with the default pseudorandom seed as the \enquote{5pAHOD} model. We let the term \enquote{5pAHOD-set} denote a corresponding choice of five parameters. The 5pAHOD-sets referenced throughout this work are listed in Table~\ref{tab:5pAHOD-sets}.

%% table -- baseline 5pAHOD-sets
\begin{table}[h]
\centering
\begin{tabular}{c c c}
\hline
Parameter & A & B \\
\hline
$\mathrm{log}M_{\mathrm{cut}}$   & $13.146692$ & 12.899689 \\
$\mathrm{log}M_1$   & $13.922568$ & 13.557676 \\
$\sigma$     & $0.554461$ & 0.062965 \\
$\alpha$     & $1.838036$ & 1.826784 \\
$\kappa$     & $0.531290$ & 2.90944 \\
\hline
\end{tabular}
\caption{\normalfont 
5pAHOD-sets that are referenced throughout this work. For example, the 5pAHOD-set in column \enquote{A} will be referenced as \enquote{5pAHOD-set-A}. Up to our filters on the 5pAHOD parameters, $\bar{n}$, and $f_\mathrm{sat}$, these 5pAHOD-sets are practically arbitrary, having been selected for our analysis by pseudorandom processes.
}
\label{tab:5pAHOD-sets}
\end{table}

The 5pAHOD model can be extended to include velocity bias \citep{guo2015velocity,Yuan_2018_vb}, assembly bias \citep{yuan2021evidence,hadzhiyska2024modest}, or satellite profile bias \citep{yuan2022abacushod}, but as discussed previously, we do not need to consider these extensions to accomplish the goals of this work.\footnote{The {\tt \textsc{AbacusHOD}} functionality is described at \url{https://abacusutils.readthedocs.io/en/latest/hod.html}. In this work we use the default parameter values for velocity bias and assembly bias; namely, $A_c=B_c=A_s=B_s=0$, $\alpha_c=0$, $\alpha_s=1$, and $s=s_v=s_p=s_r=1$.}

%%%   %%%   %%%
%%%   %%%   %%%
%%%   %%%   %%%
% Prerequisites -- Observables

\subsection{Observables}
\label{sec:prereq:obs}

\begin{figure*}[t!]
    \centering
    \includegraphics[width=\textwidth]{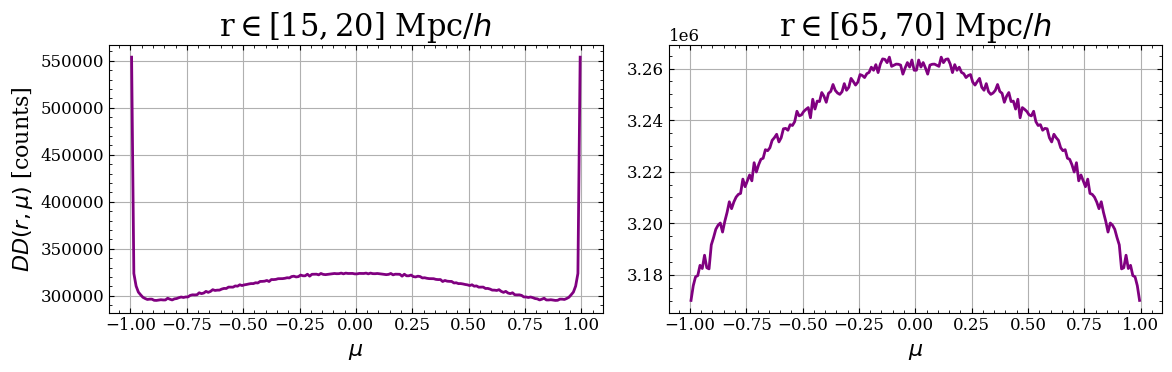}
    \caption{Two instances of $DD(r,\mu)$ for {\tt c000\_ph000}+5pAHOD-set-A (Table~\ref{tab:5pAHOD-sets}), where the left subplot corresponds to $r\in[15,20]$ Mpc/$h$ and the right subplot corresponds to $r\in[65,70]$ Mpc/$h$. One can see that small-scale clustering receives nontrivial contributions from extreme-angle galaxy pairs.}
    \label{fig:DDvmu}
\end{figure*}

For each galaxy catalog we measure the 2PCF as a function of pair separation and orientation. We denote the separation between two galaxies by $r$, and the cosine of the angle between the pair-separation vector and the line of sight by $\mu$. Galaxy pairs are binned in both quantities, with $N_r$ separation bins spanning $r_\mathrm{min}$ to $r_\mathrm{max}$ and $N_\mu$ angular bins spanning $\mu_\mathrm{min}$ to $\mu_\mathrm{max}$. The resulting pair-count histogram is denoted $DD(r,\mu)$ and is computed using the {\tt Corrfunc} algorithm \citep{Sinha_2019}. An example of $DD(r,\mu)$ in two $r$-bins is shown in Figure~\ref{fig:DDvmu}.

%% fig -- prereq -- DDvmu

We compare this measured pair count to the number expected for a random catalog with the same number density, which can be computed analytically. For a bin located at $(r',\mu')$, random pair counts can be computed as
\begin{multline}
    RR(r',\mu') = \\ 
    \bar{n} \cdot \frac{4\pi}{3}\left[ {(r'+\Delta r')}^3 - {r'}^3 \right] 
    \cdot N_{\mathrm{gal}} \cdot \frac{1}{2} 
    \cdot \frac{\mu_\mathrm{max} - \mu_\mathrm{min}}{N_\mu} \,,
\end{multline}
where $N_\mathrm{gal}$ is the number of galaxies in the catalog. The 2PCF is then defined as
\begin{equation}
    \xi(r,\mu) \equiv \frac{DD(r,\mu)}{RR(r,\mu)}-1 \,.
\end{equation}

We compress $\xi(r,\mu)$ into Legendre multipoles,
\begin{equation}
    \xi^{(l)}(r) = (2l+1) \bigl\langle P_l(\mu) \xi(r,\mu) \bigr\rangle_\mu \,,
\end{equation}
where $P_l$ is the $l^\mathrm{th}$ Legendre polynomial and $\bigl\langle X \bigr\rangle_\mu$ denotes the average of $X$ over the $\mu$ bins. In this work we use the monopole, $l=0$, and quadrupole, $l=2$, which contain the dominant 2PCF information relevant for our analysis.

\begin{figure*}[t!]
    \centering
    \includegraphics[width=\textwidth]{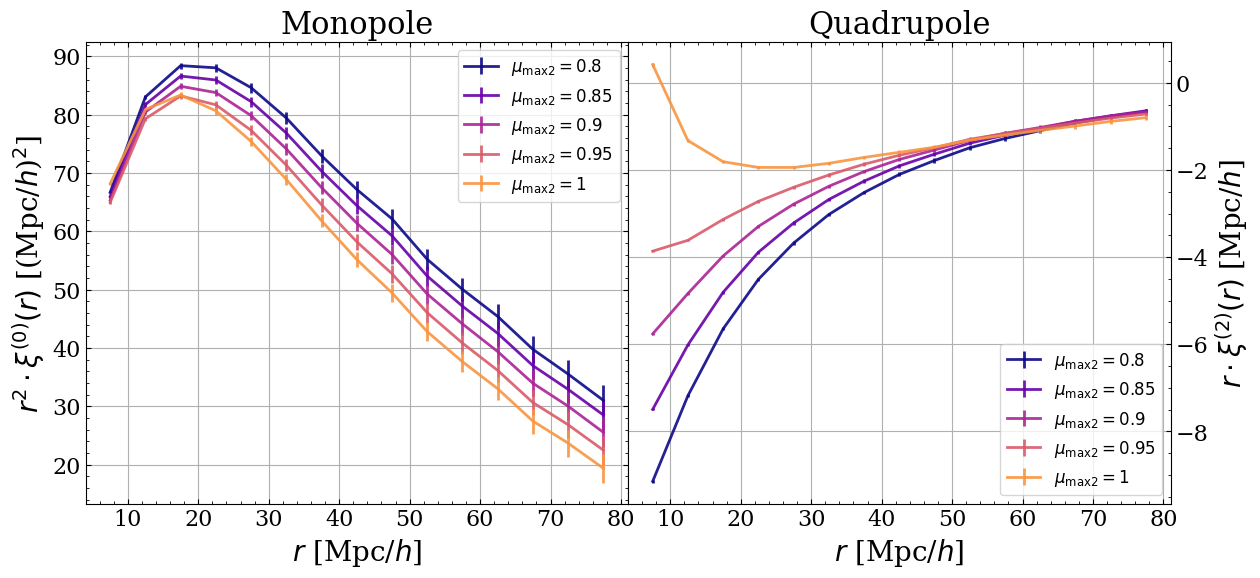}
    \caption{2PCF multipoles for several values of $\mu_\mathrm{max2}$ produced from {\tt c000\_ph000}+5pAHOD-set-A. The uncertainties shown have been recomputed for the corresponding values of $\mu_\mathrm{max2}$. In our methodology we only consider $\mu_\mathrm{max2}\in\{0.9,1.0\}$. The source causing the strong effect of $\mu_\mathrm{max2}$ in the small-scale quadrupole is the excess of extreme-angle galaxy pairs that can be seen in Figure~\ref{fig:DDvmu}.}
    \label{fig:mumax}
\end{figure*}

\begin{figure*}[t!]
    \centering
    \includegraphics[width=\textwidth]{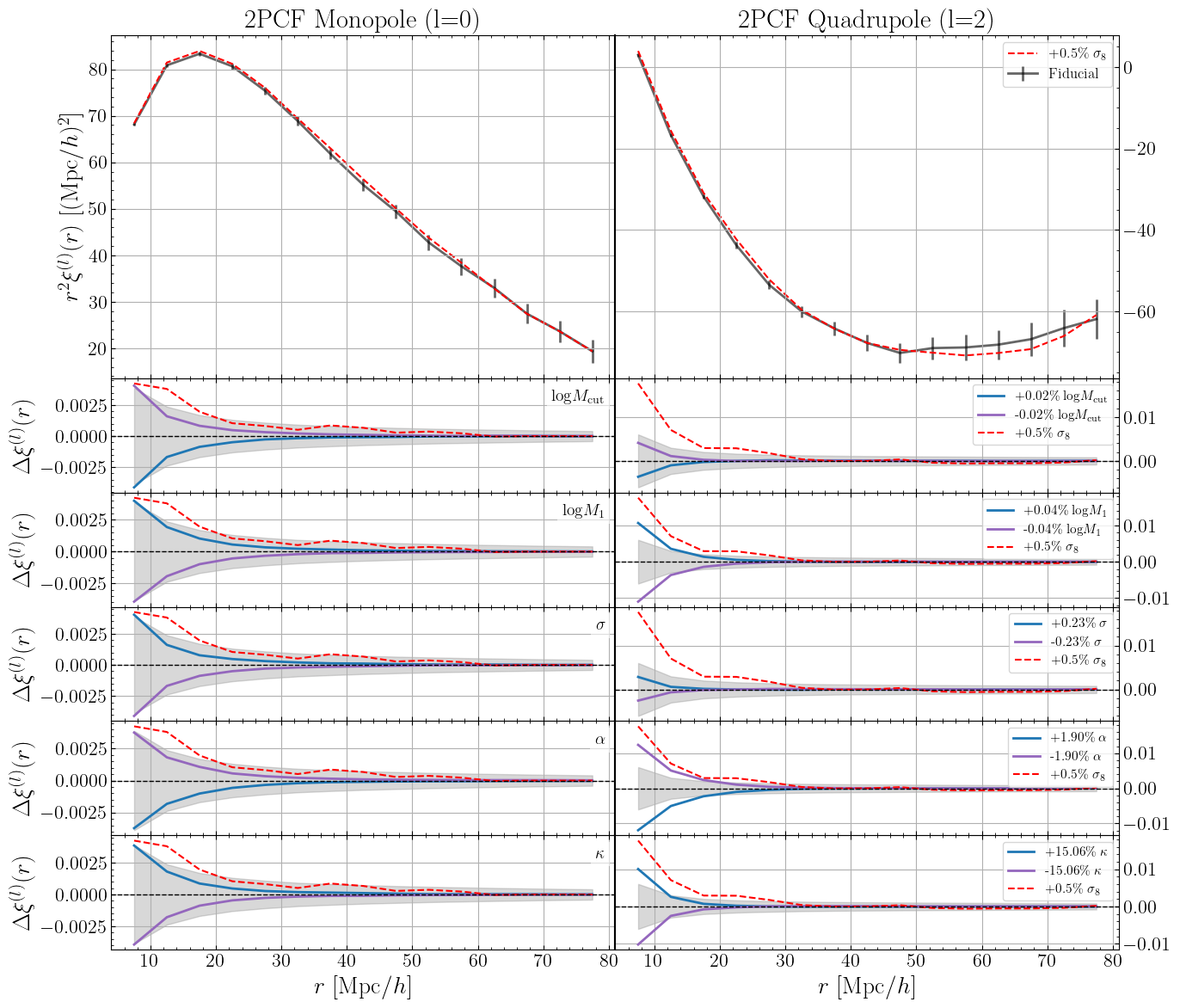}
    \caption{
    2PCF multipoles for fixed {\tt ph000}, but varying 5pAHOD and cosmological parameters about {\tt c000}+5pAHOD-set-A. 
    The top panel shows the monopole and quadrupole for {\tt c000\_ph000}+5pAHOD-set-A versus {\tt c125\_ph000}+5pAHOD-set-A. The latter is chosen because it represents a $+0.5\%$ change in $\sigma_8$, an important parameter in small-scale clustering (e.g. \citealt{leauthaud2017lensing,yuan_garrison_eisenstein_wechsler_2022,amon2023consistent}). 
    Note that while {\tt \textsc{AbacusSummit}} lists {\tt c125} as a nominal $+0.5\%$ change in $\sigma_8$, and hence our labeling, $A_s$ is covaried therein. 
    The subsequent five rows of subplots respectively show the differentials between the base case and a $\pm0.02\%$ change in $\mathrm{log}M_{\mathrm{cut}}$, a $\pm0.4\%$ change in $\mathrm{log}M_1$, a $\pm0.23\%$ change in $\sigma$, a $\pm1.90\%$ change in $\alpha$, and a $\pm15.06\%$ change in $\kappa$.
    The $\sigma_8$--$A_s$ variation of {\tt c125} is shown on these plots for comparison.
    The shaded regions depict the differential error bars corresponding to those in the first row of plots. These plots demonstrate the hierarchical sensitivity of the 5pAHOD model, especially at small scales. $\mu_\mathrm{max2}=1.0$ in all of these curves.}
    \label{fig:monoquad_HODsigma8}
\end{figure*}

The binning parameters $r_\mathrm{min}$, $r_\mathrm{max}$, $\mu_\mathrm{min}$, and $\mu_\mathrm{max}$ define the initial pair-count grid. In the statistical analysis, we may further restrict the range of bins used in the data vector. We denote by $r_\mathrm{min2}$ the minimum separation-bin lower bound included in the fit (in units of Mpc/$h$), and by $\mu_\mathrm{max2}$ the maximum value of $|\mu|$ included in the multipole calculation. 
\linebreak 
Thus, $\mu_\mathrm{max2}<1$ removes the most line-of-sight-oriented pairs, which may be less reliable in observational applications \citep{hawkins20032df,Guo_2012,Hahn_mumax,bianchi2017unbiased,mohammad2020completed}. An example of multipole sensitivity to $\mu_\mathrm{max2}$ is shown in Figure~\ref{fig:mumax}. Multipole sensitivity to cosmological and 5pAHOD parameters is shown in Figure~\ref{fig:monoquad_HODsigma8}.
%Multipole sensitivity to $\sigma_8$ and 5pAHOD parameters is shown in Figure~\ref{fig:monoquad_HODsigma8}.

%% fig -- prereq -- mumax

%% fig -- prereq -- 5pAHOD/cosmo

%%%   %%%   %%%
%%%   %%%   %%%
%%%   %%%   %%%
% Prerequisites -- Covariance
\subsection{Covariance}
\label{sec:prereq:cov}

\begin{figure}[t!]
    %\centering
    \includegraphics[width=\columnwidth]{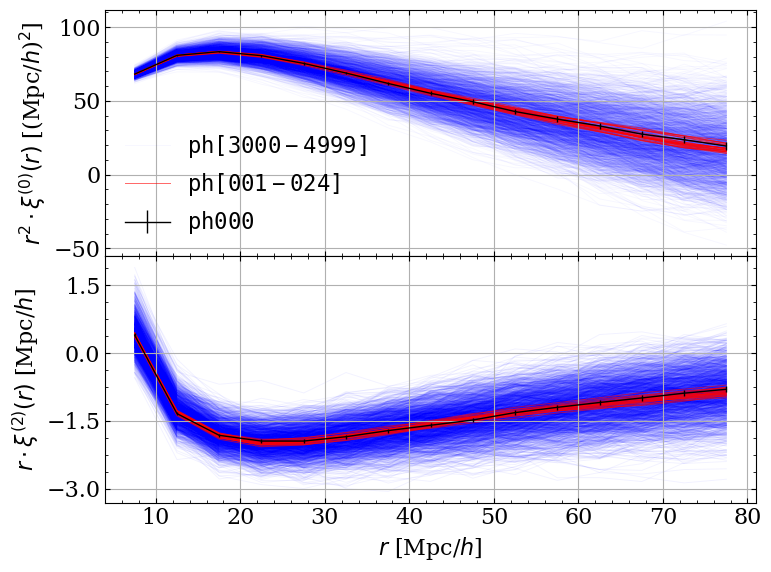}
    \caption{2PCF multipoles for fixed {\tt c000}+5pAHOD-set-A, but different phases. {\tt ph000} is shown in black. The 24 remaining cubic box phases, {\tt ph[001--024]}, are shown in red. The blue lines denote the small box phases. These are nominally listed as {\tt ph[3000--4999]}, but only 1830 of these are considered in our analysis (see Section~\ref{sec:prereq:cov}). The uncertainties shown on {\tt ph000} are taken from the covariance matrix produced from the 1830 small boxes according to the method described in Section~\ref{sec:prereq:cov}. Each curve corresponds to $\mu_\mathrm{max2}=1.0$.}
    \label{fig:smallbox_phase}
\end{figure}

\begin{figure}[t!]
    %\centering
    \includegraphics[width=\columnwidth]{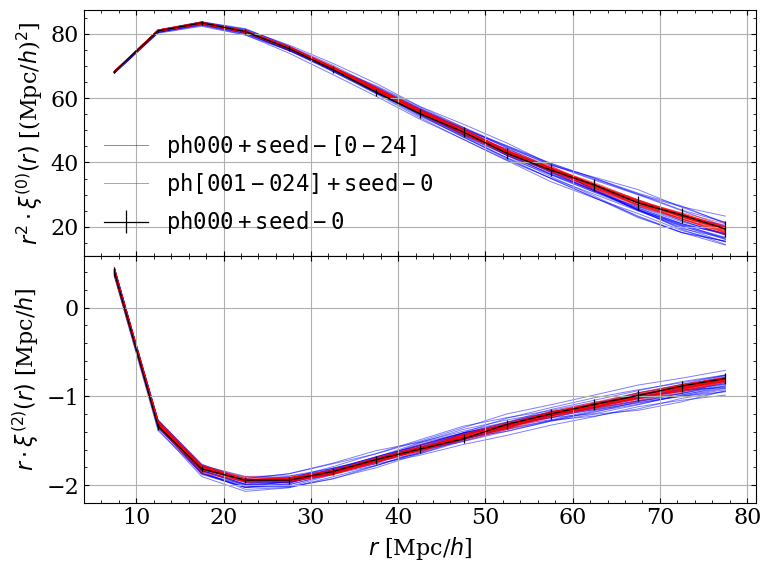}
    \caption{2PCF multipoles for fixed {\tt c000}+5pAHOD-set-A, varying phase (blue, 25 phases) and the {\tt \textsc{AbacusHOD}} seed for pseudorandom galaxy population (red, 25 seeds). The spread due to seed is smaller than the spread due to phase, but the seed spread is not negligible. We treat the seed as part of the model (fixed at the default value) and construct covariance matrices from phase variation alone (Section~\ref{sec:prereq:cov}).}
    \label{fig:phase_direct_plot}
\end{figure}

To numerically compare 2PCF multipoles, we need covariance matrices quantifying the uncertainty of each. There are two main sources of uncertainty one might consider in a simulation-based mock analysis. The first is sample variance, since different finite-volume realizations of an otherwise-fixed cosmology produce different clustering measurements (Figure~\ref{fig:smallbox_phase}). The second is the stochasticity of the galaxy--halo connection, insofar as {\tt \textsc{AbacusHOD}} assigns galaxies to halos pseudorandomly for a given 5pAHOD-set, and the resultant catalog generically depends on the {\tt \textsc{AbacusHOD}} seed (Figure~\ref{fig:phase_direct_plot}). We assume that the uncertainty on any set of 2PCF multipoles is described entirely by the phase variation,\footnote{We will address this matter in future work as well. We make this assumption, which follows common practice in the literature, strictly for computational simplicity.} thereby treating the seed-dependence as negligible.\footnote{Arguably, the seed variation \textit{is} contained within the covariance matrices produced from the 1830 small boxes: even though we kept the seed fixed at its default value in each, the 1830 small boxes have unique halo distributions, such that the \textit{effect} of using the same seed should amount to a practically unique contribution at the catalog level. However, the only way to explicitly verify this reasoning would be to test it, which we will save for future work.} Specifically, we populate 1830 {\tt c000} small boxes\footnote{53 of the 1883 small box simulations failed for technical reasons.} with the same 5pAHOD-set and compute the 2PCF multipoles for each. We then form the sample covariance of these multipole data vectors,
\begin{equation}
    C_{ij}^{\mathrm{small}}
    =
    \frac{1}{N_\mathrm{box}-1}
    \sum_{k=1}^{N_\mathrm{box}}
    \left(\xi_i^{(k)}-\bar{\xi}_i\right)
    \left(\xi_j^{(k)}-\bar{\xi}_j\right),
\end{equation}
where \(N_\mathrm{box}=1830\), \(\xi_i^{(k)}\) is the \(i^\mathrm{th}\) element of the multipole data vector measured in the \(k^\mathrm{th}\) small box, and \(\bar{\xi}_i\) is the mean over small boxes. We denote the resulting covariance matrix by \(\mathbf{C}_\mathrm{small}\), whose $(ij)^\mathrm{th}$ element is $C_{ij}^\mathrm{small}$.

Because the small boxes have side length \(500\;\mathrm{Mpc}/h\), while the cubic boxes used for the main analysis have side length \(2000\;\mathrm{Mpc}/h\), we scale the covariance by the ratio of volumes:
\begin{equation}
    \mathbf{C}_\mathrm{cubic}
    =
    \mathbf{C}_\mathrm{small}
    \left(
    \frac{500 \ \mathrm{Mpc}/h}{2000 \ \mathrm{Mpc}/h}
    \right)^3
    =
    \frac{1}{64}\mathbf{C}_\mathrm{small}.
\end{equation}
This scaling assumes that the covariance is inversely proportional to survey volume.\footnote{Note that we do not apply any smoothing to the covariance matrix. For 1830 small boxes, the Hartlap factor corresponding to our maximum number of data points (30) is 1.017, which we take as sufficiently close to 1 to not demand any corrections.} The small-box multipoles used to construct the covariance are shown in Figure~\ref{fig:smallbox_phase}.

When comparing a mock data vector to a model prediction, we use the covariance of their difference. Assuming the fluctuations in the two vectors are independent, the total covariance is the sum of the two contributions,
\begin{equation}
    \mathbf{C}_{\mathrm{tot}}
    =
    \mathbf{C}_{\mathrm{data}}
    +
    \mathbf{C}_{\mathrm{model}} .
    \label{equ:Ctot_Cdat_Cthe}
\end{equation}
When both vectors are measured from cubic boxes with the same covariance, this reduces to
\begin{equation}\label{equ:Ctot2}
    \mathbf{C}_{\mathrm{tot}}
    =
    2\mathbf{C}_{\mathrm{cubic}} .
\end{equation}
For real data, or for comparisons involving different survey volumes or mock constructions, the two covariance contributions need not be equal.

%% fig -- prereq -- phase

%% fig -- prereq -- seed

%%%   %%%   %%%
%%%   %%%   %%%
%%%   %%%   %%%
% Prerequisites -- Statistical Analysis

\subsection{Statistical analysis}
\label{sec:prereq:stat}

Given a measured 2PCF multipole data vector, $\bm{\xi}_\mathrm{data}$, and a model prediction, $\bm{\xi}_\mathrm{mod}$, we define
\begin{equation}
    \Delta \bm{\xi}
    \equiv
    \bm{\xi}_\mathrm{data}
    -
    \bm{\xi}_\mathrm{mod}.
\end{equation}
Using the total covariance matrix $\mathbf{C}_\mathrm{tot}$ defined in Equation~\eqref{equ:Ctot_Cdat_Cthe}, the goodness of fit is measured by
\begin{equation}
    \chi^2
    =
    \Delta \bm{\xi}^{T}
    \mathbf{C}_\mathrm{tot}^{-1}
    \Delta \bm{\xi}.
    \label{equ:chisq}
\end{equation}
We report most of our results in terms of the reduced chi-squared,
\begin{equation}
    \chi_\nu^2
    \equiv
    \frac{\chi^2}{\nu},
\end{equation}
where $\nu$ is the number of degrees of freedom, equal to the number of data points minus the number of fitted parameters.

For most of the analyses in this paper, the data vector contains 30 2PCF multipole bins and the fitted nuisance parameters are the five parameters of the 5pAHOD model, which would trivially suggest taking $\nu=25$ in our $\chi^2_\nu$ computations.\footnote{Considering the manifest hierarchical dependence of clustering observables to the 5pAHOD parameters, it is not clear that each parameter contributes equally to the degrees-of-freedom computation. We will assume this to be the case without proof, and save a further investigation of this matter for future work. For papers which address similar questions, see: \citet{spiegelhalter1998bayesian,kunz2006measuring,gelman2014understanding,raveri2019concordance,handley2019quantifying}} Assuming a Gaussian likelihood and $\nu=25$, the 68.3\%, 95.5\%, and 99.7\% upper-tail thresholds correspond to
\begin{equation}
    \chi_\nu^2 = 1.11,\quad 1.52,\quad 1.95,
\end{equation}
respectively. We use these thresholds as convenient reference values for interpreting the goodness of fit. Under the Gaussian assumption, if some model is true, then the probability that it yields a $\chi^2_\nu$ at or above one of these thresholds is 31.7\%, 4.5\%, or 0.3\%, respectively.

In principle, one could use the same statistic to fit a continuous cosmological parameter space by minimizing over both cosmological and HOD parameters or otherwise applying some interpolation scheme. However, we do not attempt this. In preparation, we attempted an interpolation over the 81-cosmology grid, but found that results were highly sensitive to the underlying assumptions of the interpolator employed in any given case. The most conservative choice to avoid this complication, as well as that of validating the interpolator to keep with our primary goal, is to avoid cosmology interpolation altogether. 
Instead, we evaluate the goodness of fit separately at each available cosmology through profile likelihood of the HOD parameters.

%%%   %%%   %%%
%%%   %%%   %%%
%%%   %%%   %%%
\subsection{Interpolated HOD predictions}
\label{sec:prereq:interp}

A direct 2PCF prediction is obtained by populating a specified {\tt \textsc{AbacusSummit}} halo catalog with {\tt \textsc{AbacusHOD}}, measuring pair counts with {\tt Corrfunc}, and computing the Legendre multipoles. A single direct computation takes on the order of several seconds, making it impractical to use direct evaluations for all minimization, validation, and mock-analysis steps. To expedite minimization, for each cosmology we construct a unique 2PCF multipole interpolator in HOD parameter space by directly computing 2PCF multipoles with dense pseudorandom sampling of 5pAHOD-sets (Section~\ref{sec:Methods:catalogs_interp}). We refer to the corresponding interpolated five-parameter model as the \enquote{5pIHOD model}, and a corresponding set of five parameters as a \enquote{5pIHOD-set}. 

We do not require the 5pIHOD model to reproduce the 2PCF multipoles of the 5pAHOD model when 5pIHOD-set = 5pAHOD-set. Since we are profiling rather than marginalizing, it suffices that some 5pIHOD-set exists that reproduces the multipoles of any possible 5pAHOD-set, even if 5pIHOD-set $\neq$ 5pAHOD-set, provided this 5pIHOD-set can be located by our global optimizer (see validation procedure in Section~\ref{sec:Methods:Val}). In this sense, we are not interested in the particular values of the HOD parameters, but simply in locating the HOD parametrization that yields the best-fit between model and data. 

We conclude this Section by summarizing our terminology pertaining to the galaxy--halo connection: the \enquote{HOD model} is the general term for any model that instantiates a galaxy--halo connection; the \enquote{5pHOD model} is the HOD model described in Equations~(\ref{equ:Ncen}--\ref{equ:Nsat}); the \enquote{5pAHOD model} is the {\tt \textsc{AbacusHOD}} implementation of the 5pHOD model which differs due to algorithmic choices for efficiency; the \enquote{5pIHOD model} denotes our interpolated version of the 5pAHOD model.

%%%   %%%   %%%
%%%   %%%   %%%
%%%   %%%   %%%
% Section: HODmin
%%%   %%%   %%%
%%%   %%%   %%%
%%%   %%%   %%%
% HODmin

\begin{samepage}
\section{\normalsize{\tt HOD\MakeLowercase{min}} \footnotesize{Algorithm}}
\label{sec:HODmin}

To obtain our profiled $\chi^2$ values, we require a reliable global optimization algorithm. However, the $\chi^2$ surface is non-smooth, contains many local minima, and varies by several orders of magnitude across the allowed parameter space. These difficulties pose a challenge even in five dimensions. A purely local optimizer may converge to a non-global local minimum, whereas purely pseudorandom sampling is inefficient. To address this dilemma, we developed {\tt HODmin}, a two-stage global optimization algorithm that combines the advantages of local minimization and pseudorandom sampling to obtain profiled $\chi^2$ values. The first stage uses pseudorandom sampling to locate a promising region of 5pIHOD space, and the second stage applies a local optimizer to refine the minimum within that region.
\end{samepage}

{\tt HODmin} begins with broad bounds on each of the 5pIHOD parameters, which can be seen in Table~\ref{tab:initHODbounds}. During the first stage, denoted \enquote{{\tt HODmin}-S1}, these bounds are compressed through ten successive sampling substages. In each substage, trial 5pIHOD-sets are drawn pseudorandomly from within the current bounds. Each trial set is first required to satisfy a number-density filter: its predicted number density must lie within $\pm20\%$ of the target value $\bar{n}=0.0005$ Mpc/$h$.\footnote{This is the DESI-DR3 target number density for LRGs \citep{zhou2023desi}.} Points that fail this cut are discarded without any subsequent computation. For the accepted points, the 2PCF multipoles are evaluated and compared to the data vector; with the appropriate covariance matrices, $\chi^2$ is computed according to Equation~\eqref{equ:chisq}.

Each substage continues until a predetermined number of successful samples has been collected. We use $10^3$ successful samples in the first substage, $10^4$ in the second substage, and $10^3$ in each of the remaining eight substages. The successful samples are then sorted by $\chi^2$. The best-performing part of this sorted list is used to define the parameter bounds for the next substage. Depending on the parameter and substage, the next bounds are either left unchanged, set to the minimum and maximum values among the selected low-$\chi^2$ samples, or fixed to the median of those samples for the remainder of {\tt HODmin}-S1. This progressive compression allows the algorithm to discard poorly-fitting regions while continuing to search broadly within the most viable regions. 

At the end of every {\tt HODmin}-S1 substage, we record the best $\chi^2$ value found and the corresponding 5pIHOD-set. After all ten substages are complete, {\tt HODmin}-S1 returns the best of these ten parameter sets. This point is then used as the initial point for a local search in the second stage, denoted \enquote{{\tt HODmin}-S2}. This stage uses the {\tt HODmin}-S1 output as the initial point for a Nelder--Mead minimization of the same $\chi^2$ function. The purpose of {\tt HODmin}-S1 is to provide a starting point in a low-$\chi^2$ region, making the subsequent local optimization less likely to converge to an irrelevant local minimum. The final output of {\tt HODmin} is the Nelder--Mead minimum from {\tt HODmin}-S2, together with its best-fitting 5pIHOD-set. We also retain the best {\tt HODmin}-S1 $\chi^2$ value as a diagnostic to examine the extent of improvement from local refinement.

Although our implementation is tailored to the 5pIHOD model and the $\chi^2$ statistic, the logic of {\tt HODmin} is more general, being most useful for objective functions that are expensive to evaluate, lack reliable gradients, and contain many local minima. In fact, we originally developed {\tt HODmin} with the intention of validating a global optimizer for the 5pAHOD model directly, for which an individual computation takes on the order of $5-10$ seconds (cf. \citealt{yuan2021evidence}), whereas the 5pIHOD model interpolator computations are relatively instantaneous. In light of our rapid objective function, a different global optimizer certainly could be employed and validated instead of {\tt HODmin}. Rather, we use {\tt HODmin} in this work to demonstrate its functionality. {\tt HODmin} is similar in spirit to basin-hopping and other global-plus-local optimization strategies, but its first-stage compression procedure is tuned to the structure of 5pHOD-based $\chi^2$ surfaces. Our general implementation of {\tt HODmin} is described in Section~\ref{sec:Methods:Procedure}, its convergence validation is given in Section~\ref{sec:Methods:Val}, and our particular implementations are listed in Section~\ref{sec:Methods:Tests}.

\begin{table}[h]
\centering
\begin{tabular}{c c}
\hline
Parameter & Initial Bounds \\
\hline
$\mathrm{log}M_{\mathrm{cut}}$   & $[12,\,14]$ \\
$\mathrm{log}M_1$   & $[\mathrm{log}M_{\mathrm{cut}},\,15]$ \\
$\sigma$     & $[0,\,1]$ \\
$\alpha$     & $[0,\,2]$ \\
$\kappa$     & $[0,\,3]$ \\
\hline
\end{tabular}
\caption{\normalfont Initial bounds for 5pHOD parameters.}
\label{tab:initHODbounds}
\end{table}

%%%   %%%   %%%
%%%   %%%   %%%
%%%   %%%   %%%
% Section: Methods
%%%   %%%   %%%
%%%   %%%   %%%
%%%   %%%   %%%
% Methods
\section{Methods}\label{sec:Methods}

In this Section we describe the pipeline used to produce our mock-analysis results. We first construct the ingredients needed for each test: direct 5pAHOD galaxy catalogs and interpolators in 5pIHOD parameter space (Section~\ref{sec:Methods:catalogs_interp}), followed by mock data vectors and covariance matrices (Section~\ref{sec:Methods:Repo:Mocks}). Next, we define a single analysis configuration and explain how {\tt HODmin} is run on that configuration to obtain a profiled minimum $\chi^2$ (Section~\ref{sec:Methods:Procedure}). We then validate the convergence of {\tt HODmin} using self-consistent mocks for which the expected minimum is known to be near zero (Section~\ref{sec:Methods:Val}). Finally, we list the primary and secondary tests used to measure the floor and ceiling constraints presented in Section~\ref{sec:results} (Section~\ref{sec:Methods:Tests}).

%%%   %%%   %%%
%%%   %%%   %%%
%%%   %%%   %%%
% Methods -- Catalogs and interpolators
\subsection{Galaxy catalogs and interpolators}
\label{sec:Methods:catalogs_interp}

For each of the 81 cosmologies tested in our analysis (see Table~\ref{tab:HODmin_fin}), we construct training catalogs from the respective {\tt ph000} cubic box. We employ {\tt \textsc{AbacusHOD}} to populate the halo catalog for 5pAHOD-sets (see description in Section~\ref{sec:prereq:HOD}). The 5pAHOD-sets are drawn pseudorandomly from the initial bounds in Table~\ref{tab:initHODbounds}, with the seed fixed to its default value throughout.

\begin{figure*}[t!]
    \centering
    \includegraphics[width=\textwidth]{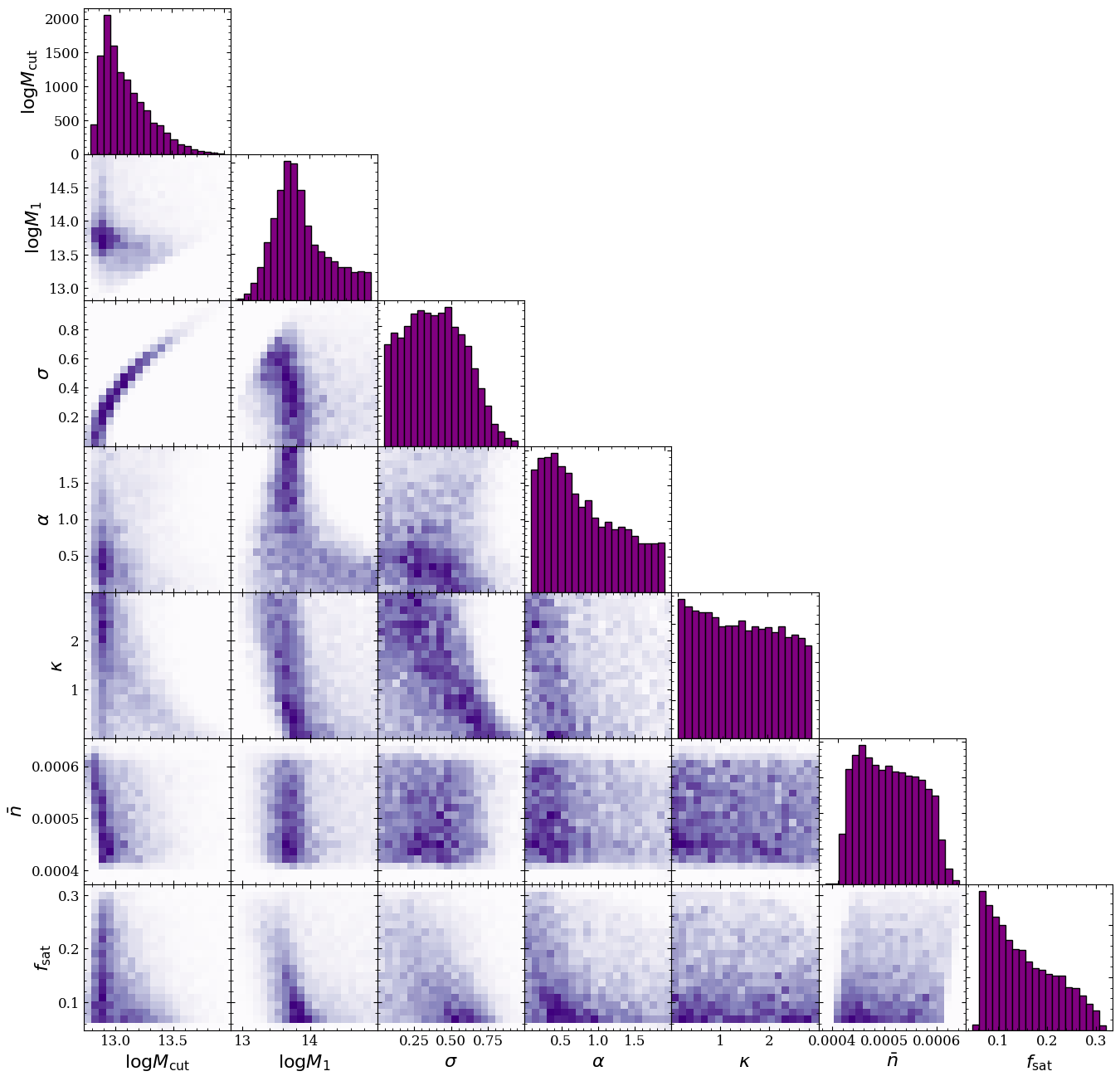}
    \caption{Sampling density of the filtered interpolation data vectors for {\tt c000\_ph000}. The nontrivial shapes arise from the filters on $\bar{n}$ and $f_\mathrm{sat}$; in the unfiltered case the distribution against the 5pAHOD parameters would be approximately uniform. The 500 unfiltered sampling vectors are omitted. Note that some $\bar{n}$ and $f_\mathrm{sat}$ values fall outside the nominal cuts because our filtration acts on a far quicker but slightly less accurate computation of $\bar{n}$ and $f_\mathrm{sat}$ offered in the {\tt \textsc{AbacusHOD}} functionality that precedes the more expensive population of galaxies.}
    \label{fig:interp_sampdens}
\end{figure*}

We retain catalogs whose mean galaxy number density lies within $\pm 20\%$ of $\bar{n}_{\rm cen}=0.0005$ Mpc/$h$ and whose satellite fraction lies in the range $f_\mathrm{sat}\in[0.05,0.25]$. These broad cuts restrict the training set to LRG-like catalogs (cf. \citealt{zhou2023desi}) while leaving substantial freedom in the HOD parameters otherwise. For each cosmology, we retain 12{,}000 catalogs satisfying these cuts. We also generate 500 additional catalogs drawn from the same initial HOD bounds but without applying the $\bar{n}$ or $f_\mathrm{sat}$ filters. These extra points improve the behavior of the interpolator near the boundary of the sampled region and reduce the frequency of invalid, or {\tt nan}, interpolator returns. The resulting sampling density of the filtered training set is shown in Figure~\ref{fig:interp_sampdens}. The nontrivial shapes in 5pAHOD--$\bar{n}$--$f_\mathrm{sat}$ space arise from the filters on $\bar{n}$ and $f_\mathrm{sat}$. If the filters had not been implemented, we would expect an approximately uniform sampling distribution. The notable $\mathrm{log}M_{\mathrm{cut}}$--$\sigma$ contour arises almost entirely from the $\bar{n}$ constraint, whereas the nontrivial shapes involving $\mathrm{log}M_1$, $\alpha$, and $\kappa$ arise from both the $\bar{n}$ and $f_\mathrm{sat}$ filters. Low values of $\mathrm{log}M_1$ are largely suppressed, and high values only tend to occur for low $\alpha$. We also observe that $\kappa$ is the least sensitive parameter to our filters.

From each of the 12{,}500 catalogs, we compute 2PCF multipoles following Section~\ref{sec:prereq:obs}. We consider two values of the angular cut, $\mu_\mathrm{max2}\in\{0.9,1.0\}$, and construct a separate multipole set for each. We construct a piecewise-linear interpolator in the 5pIHOD space using 
{\tt scipy.interpolate.LinearNDInterpolator} for each cosmology and each value of $\mu_\mathrm{max2}$. A number-density interpolator is trained analogously for each cosmology. 

\begin{figure*}[t!]
    \centering
    \includegraphics[width=\textwidth]{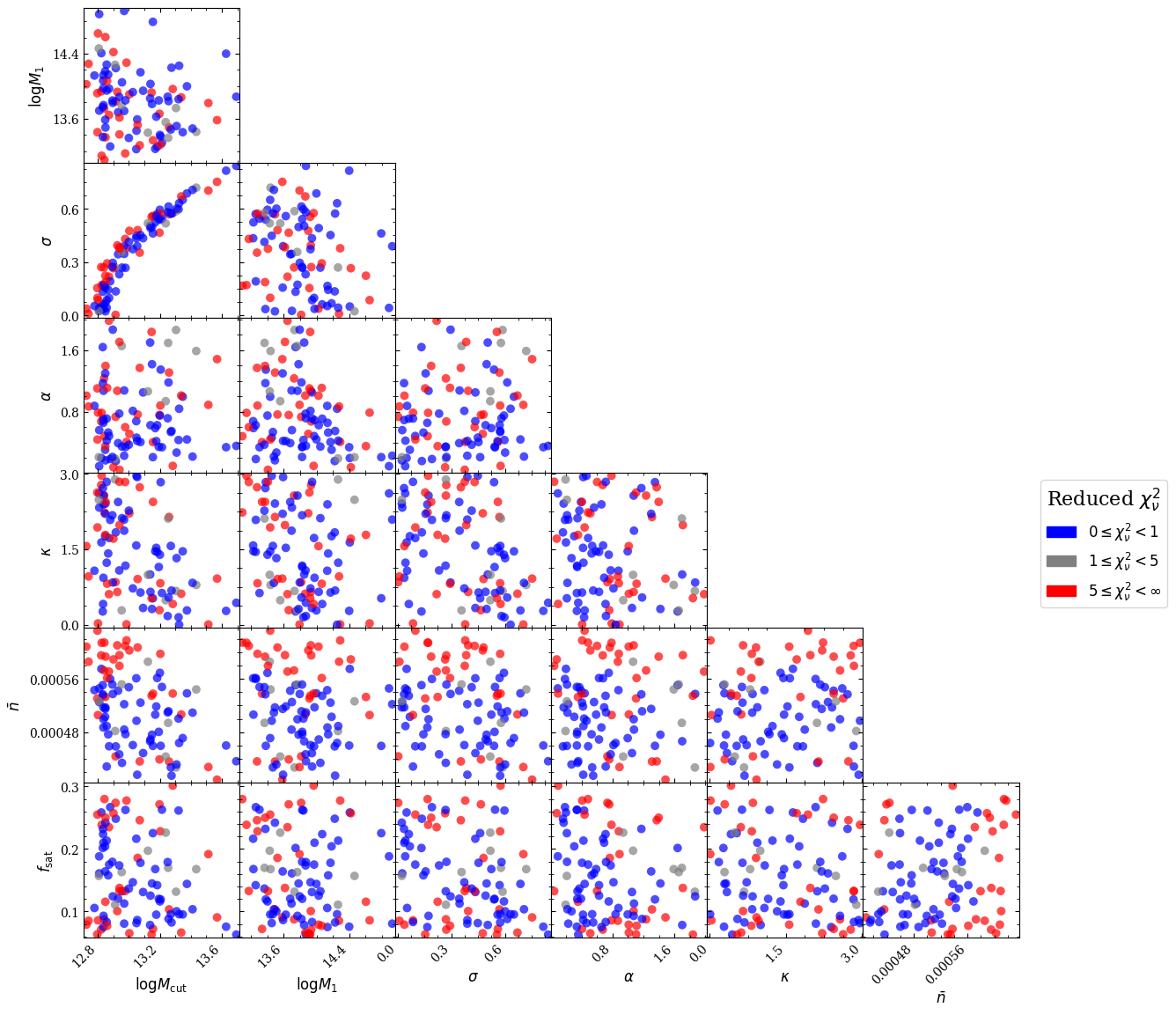}
    \caption{Accuracy of the 2PCF multipole interpolator (5pIHOD) relative to {\tt \textsc{AbacusHOD}} (5pAHOD) for {\tt c000\_ph000}, where for each case, 5pAHOD-set = 5pIHOD-set. These pseudorandom 5pAHOD-sets were generated approximately within our bounds on HOD parameters, $\bar{n}$, and $f_\mathrm{sat}$ (see comment on Figure~\ref{fig:interp_sampdens} regarding outliers). $\chi^2_\nu$ is our proxy for comparison and is computed directly without any profiling. Good agreement ($\chi^2_\nu \lesssim 1$) is common but not reliable; outliers tend to lie near the edges of the sampled $\bar{n}$--$f_\mathrm{sat}$ region. This motivates the tighter $\bar{n}$ and $f_\mathrm{sat}$ ranges used for validation and test mocks.}
    \label{fig:interp_v_AHOD}
\end{figure*}

The resulting interpolators define the 5pIHOD model for which we employ {\tt HODmin}. These interpolators are not exact reproductions of the direct {\tt \textsc{AbacusHOD}} evaluations for fixed 5pHOD-sets, as shown in Figure~\ref{fig:interp_v_AHOD}, where we show a {\tt c000\_ph000} test that compares our interpolator to 100 pseudorandomly-generated {\tt \textsc{AbacusHOD}} mock data vectors within roughly our filters listed in Section~\ref{sec:Methods:Repo:Mocks}. The discrepancy arises from interpolation error and the pseudorandom nature of galaxy population. This does not undermine the profile-likelihood analysis, provided that the interpolated HOD space contains 2PCF vectors close enough to the relevant direct {\tt \textsc{AbacusHOD}} targets, which we explain in our validation procedure in Section~\ref{sec:Methods:Val}.

%% interp sampling

% interp v AHOD accuracy

\subsection{Mock data vectors and covariance matrices}
\label{sec:Methods:Repo:Mocks}

For each of the 81 cosmologies, we produce five mock data vectors used in the self-consistency validation described in Section~\ref{sec:Methods:Val}. Each mock data vector is generated directly from an {\tt \textsc{AbacusHOD}} catalog and consists of the 2PCF multipoles defined in Section~\ref{sec:prereq:obs}. The corresponding 5pAHOD-sets are drawn using tighter cuts than those used for the interpolator training set: $\bar{n}$ must lie within $\pm 10\%$ of $\bar{n}_\mathrm{cen}$, and the satellite fraction must satisfy $f_\mathrm{sat} \in [0.06,0.20]$. These tighter cuts keep the mock data vectors well inside the region where the interpolators are most accurate, as illustrated in Figure~\ref{fig:interp_v_AHOD}. A separate set of mock data vectors is produced for each value of $\mu_\mathrm{max2}$.

We also produce mock data vectors for each of the 25 cubic-box phases of {\tt c000}, all evaluated at 5pAHOD-set-A (Table~\ref{tab:5pAHOD-sets}). These mocks are used to assess how the profiled constraints vary with phase for an otherwise fixed cosmology and galaxy--halo connection.

Finally, we compute covariance matrices using the 1830 small boxes following the procedure described in Section~\ref{sec:prereq:cov}. We construct two versions, one in which the small boxes are populated with 5pAHOD-set-A and one in which they are populated with 5pAHOD-set-B. Both covariance matrices are produced separately for each value of $\mu_\mathrm{max2}$.

%%%   %%%   %%%
%%%   %%%   %%%
%%%   %%%   %%%
% Methods -- Procedure
\subsection{Procedure}
\label{sec:Methods:Procedure}

A configuration of our experiment consists of the following: a cosmology (selecting an interpolator, with phase {\tt ph000} and the 5pIHOD model implied, but not any particular 5pIHOD-set); a mock data vector (specifying the target multipoles associated with a cosmology, phase, and 5pAHOD-set); a covariance matrix (used for both the model and data vectors in each $\chi^2$ computation, see Section~\ref{sec:Methods:Val} for justification); a scale cut $r_\mathrm{min2}$; a $\mu$-cut $\mu_\mathrm{max2}$; and a multipole choice (monopole, quadrupole, or both). For each configuration we run {\tt HODmin} 1000 times\footnote{The \textit{only} difference between each of these 1000 runs is different placement in the pseudorandom number chain.} and record the run that achieves the absolute minimum $\chi^2$. The result reported for that configuration is this minimum $\chi^2$ and its corresponding 5pIHOD-set.

The interpretation of our results rests on a small number of assumptions, which we state here explicitly. We assume that any {\tt \textsc{AbacusSummit}} simulation produces the halo distribution corresponding to its cosmology and phase with negligible error. We assume that the 5pAHOD model has sufficient flexibility to describe the exact galaxy--halo connection that follows from any cosmology and phase under consideration via some 5pAHOD-set within the bounds of Table~\ref{tab:initHODbounds} and the filters on $\bar{n}$ and $f_\mathrm{sat}$ mentioned in Section~\ref{sec:Methods:Repo:Mocks}. We assume the uncertainty on any multipole or data vector is fully described by the relevant covariance matrix, and that uncertainties between interpolation and data vectors are uncorrelated with respect to one another so that Equation~\eqref{equ:Ctot2} applies. The convergence of {\tt HODmin} to the global minimum is not guaranteed in principle; we assume the validation described in Section~\ref{sec:Methods:Val} suffices for the cases of interest.

%%%   %%%   %%%
%%%   %%%   %%%
%%%   %%%   %%%
% Methods -- Validation
\subsection{Validation}
\label{sec:Methods:Val}

We validate the convergence of {\tt HODmin} using self-consistent cases in which the expected outcome is known. In these tests, the mock data vector and the interpolator are constructed from the same cosmology and phase. Therefore, after profiling over the HOD parameters, the minimum $\chi^2$ should be close to zero, up to interpolation error and statistical noise. A failure to find such a low value would indicate either that the interpolated 5pIHOD space does not adequately represent the corresponding direct 5pAHOD mock, or that {\tt HODmin} has failed to locate the relevant minimum.

For each of the 81 cosmologies, we run this self-consistency test on five independent mock 5pAHOD-sets, giving 405 validation cases in total. The mock 5pAHOD-sets are drawn using the cuts described in Section~\ref{sec:Methods:Repo:Mocks}. We run each case with $r_\mathrm{min2}=5$, $\mu_\mathrm{max2}=1.0$, and monopole+quadrupole. This is the most restrictive configuration considered here, since it includes the largest number of data points and therefore provides the strongest test of the minimization.

For the $\chi^2$ calculations, we use a single covariance matrix throughout: the 5pAHOD-set-A covariance from Section~\ref{sec:Methods:Repo:Mocks}. {\tt \textsc{AbacusSummit}} provides small boxes only for {\tt c000}, such that per-cosmology covariance matrices are not available; we therefore use the {\tt c000}-based covariance for all cosmologies. We also use the 5pAHOD-set-A covariance for all validation mocks, even though the mocks are generated from different 5pAHOD-sets. This is an approximation, but it does not affect the purpose of the validation. The goal is to determine whether {\tt HODmin} can find a near-zero residual when the data vector and interpolator are self-consistent. In that regime, the precise covariance normalization is not the limiting factor.\footnote{It seems reasonable to expect that the covariance is an overestimate in some cases and an underestimate in some cases; if, therefore, we find that all validation cases converge to $\chi^2_\nu \ll 1$, we can be confident that our {\tt HODmin} validation withstands this covariance limitation.}

\begin{figure}[t!]
    \includegraphics[width=\columnwidth]{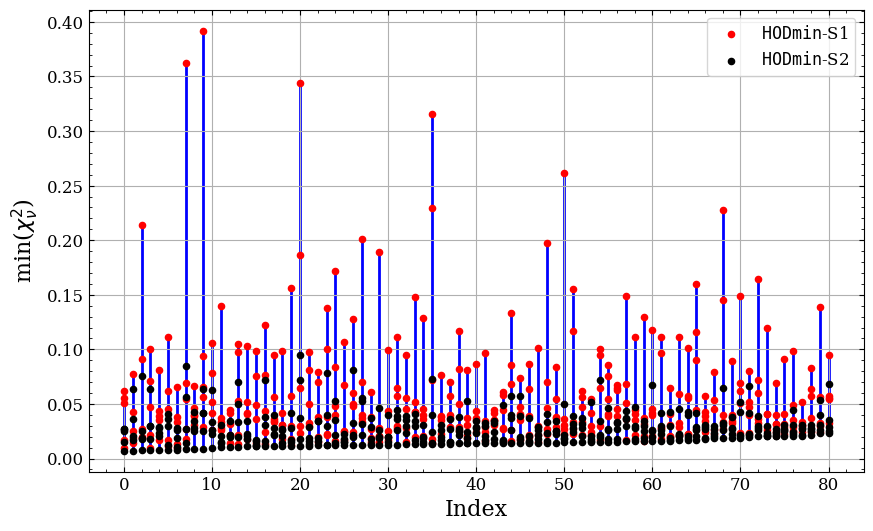}
    \caption{Validation results for each cosmology tested in the suite. For each cosmology, the best minimum $\chi^2_\nu$ values from {\tt HODmin}-S1 and {\tt HODmin}-S2 out of the 1000 independent values for all 5 validation cases are shown, such that there are 10 points per cosmology, 405 unique validation mocks across the suite, and 810 data points shown in total. \enquote{Index} represents the 81 cosmologies tested, but these are not ordered according to the {\tt \textsc{AbacusSummit}} labels. Rather, the cosmologies are ordered along the Index by increasing value of the minimum {\tt HODmin}-S2 $\chi^2_\nu$ value of the five per cosmology. Note that for any given cosmology and validation case, the minimum $\chi^2_\nu$ values plotted from {\tt HODmin}-S1 and {\tt HODmin}-S2 do not necessarily correspond to the same individual run of the {\tt HODmin} algorithm.
    We have shown the {\tt HODmin}-S1 results to demonstrate the success of our progressive domain compression prior to local optimization, but all main results in this work come from {\tt HODmin}-S2.
    }
    \label{fig:TCval_suite}
\end{figure}

The results are shown in Figure~\ref{fig:TCval_suite}. All 405 validation cases converge to $\chi^2_\nu < 0.1$, satisfying our criterion that the self-consistent fits should have $\chi^2_\nu \ll 1$. Following our validation assumption in Section~\ref{sec:Methods:Procedure}, we therefore take this test as an indication that, for the configurations used in this paper, {\tt HODmin} reliably finds the relevant low-$\chi^2$ solution when such a solution exists. The remaining science tests apply the same minimization procedure to cases where the best achievable $\chi^2$ may be nonzero because of genuine cosmological mismatch rather than optimizer failure.

%% validation linear plot

%%%   %%%   %%%
%%%   %%%   %%%
%%%   %%%   %%%
% Methods -- Tests

\subsection{Tests}
\label{sec:Methods:Tests}

We now describe the tests run in this paper. All floor measurements use the profiled minimum found by {\tt HODmin}, whose convergence was validated in Section~\ref{sec:Methods:Val}. We report results in terms of $\chi^2_\nu$ so that configurations with different numbers of data points can be compared approximately.

The first set of
tests measure the conservative floor on cosmological constraining power. We compare each of the 81 cosmologies to the mock data vector generated from {\tt c000\_ph000}+5pAHOD-set-A (Table~\ref{tab:5pAHOD-sets}). For each cosmology, we run independent tests for every combination of $r_\mathrm{min2} \in \{5,20\}$, $\mu_\mathrm{max2} \in \{0.9,1.0\}$, and multipole choice: monopole only, quadrupole only, or monopole+quadrupole. These tests determine which cosmologies can be ruled out after allowing the HOD parameters to vary freely within the broad prior bounds, and they show how the floor depends on scale cuts, multipole choice, and mock choice. 
These tests yield our secondary results.

We then run four additional tests, each modifying one aspect of the primary configuration, which we choose to be $r_\mathrm{min2}=5$, $\mu_\mathrm
{max2}=0.9$, monopole+quadrupole, and a mock data vector generated from {\tt c000\_ph000}+5pAHOD-set-A. In each of these tests, the covariance matrix used in the $\chi^2$ calculation is the {\tt c000}-based covariance matrix corresponding to the 5pAHOD-set of the mock data vector (Section~\ref{sec:Methods:Repo:Mocks}). This covariance is only intended to be exact for mocks drawn from {\tt c000} with the corresponding 5pAHOD-set; for other cosmologies this is an approximation. Since covariance matrices for the full emulator grid are not available, we use this fixed-covariance treatment throughout. 
Among the
main conclusions of the paper
are the
large relative changes in $\chi^2_\nu$
across the various tests, especially the floor--ceiling gap, and are therefore not expected to be sensitive to modest covariance differences across cosmology. For individual strongly excluded cases, however, the precise numerical value of $\chi^2_\nu$ should be interpreted at the order-of-magnitude level.\footnote{An additional reason to motivate using the covariance matrix corresponding to the data vector for all interpolation vectors, if one at least admits its applicability to the data vector, is that we are only interested in the final $\chi^2$ output by {\tt HODmin}. Via global optimization, this final $\chi^2$ corresponds to an interpolation vector that should look sufficiently similar to the data vector that a common, and hence accurate, covariance matrix is plausible. This reasoning, however, only applies in the floor tests.} The four additional tests are given as follows, where the first three tests contribute secondary results, and the final test contributes our primary result:

\textit{Phase variation across cosmologies.} We rerun the 81-cosmology suite against a mock from {\tt c000\_ph002}+5pAHOD-set-A. This test isolates the effect of changing the phase realization of the mock data vector while keeping the cosmology and galaxy--halo connection fixed otherwise.

\textit{HOD variation.} We rerun the 81-cosmology suite against a mock from {\tt c000\_ph000}+5pAHOD-set-B, using the 5pAHOD-set-B covariance matrix. This test isolates the effect of changing the HOD model used to generate the mock data vector and covariance matrix.

\textit{Within-cosmology phase scan.} We run the {\tt c000\_ph000} interpolator against 25 mock data vectors corresponding to {\tt c000\_ph[000-024]}+5pAHOD-set-A. This test characterizes the sensitivity of the profiled minimum to phase variation within a single otherwise-fixed cosmology.

\textit{Ceiling case.} As a counterpoint to the profiled floor measurements, we compute a ceiling case in which the HOD parameters are assumed known. For every cosmology in the suite, we directly compute the 2PCF multipoles for {\tt ph000}+5pAHOD-set-A and compare them to the {\tt c000\_ph000}+5pAHOD-set-A mock without minimizing over HOD parameters or interpolating. This gives the strongest constraining power available under the supposition that the exact HOD parameters are known.

%%%   %%%   %%%
%%%   %%%   %%%
%%%   %%%   %%%
% Section: Results
\section{Results}
\label{sec:results}

We present the results of the tests described in Section~\ref{sec:Methods:Tests}. The
primary
result is the comparison between the conservative floor and upper-limiting ceiling constraints in Section~\ref{sec:results:floor_ceiling}. Before presenting that comparison, we first examine several diagnostic tests 
and secondary results
that show how the profiled minimum depends on convergence, scale cuts, angular cuts, multipole choice, phase, and HOD realization. Unless otherwise stated, results are reported in terms of $\chi^2_\nu$ to allow approximate comparison between configurations with different numbers of degrees of freedom. When this comparison is not direct, for example between different $r_\mathrm{min2}$ values or different multipole choices, we note the limitation explicitly.

\subsection{Self-consistent baseline}
\label{sec:results:baseline}

We begin with a self-consistent case in which the interpolator and mock data vector are constructed from the same cosmology. Specifically, we apply the {\tt c000\_ph000}+5pIHOD interpolator to the {\tt c000\_ph000}+5pAHOD-set-A mock data vector, using $r_\mathrm{min2}=5$, $\mu_\mathrm{max2}=0.9$, and monopole+quadrupole. This case is closely related to the validation tests in Section~\ref{sec:Methods:Val}, and illustrates the behavior of {\tt HODmin} when the expected minimum is near $\chi^2=0$. This case, moreover, is the baseline upon which the subsequent tests vary.

\begin{figure}[t!]
    \includegraphics[width=\columnwidth]{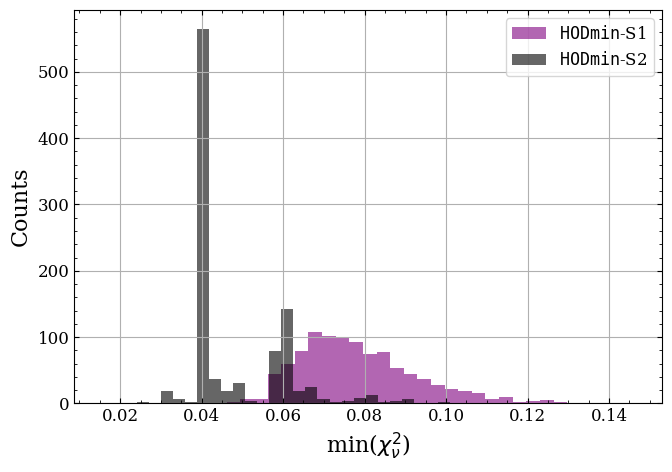}
    \caption{Distribution of minimum $\chi^2_\nu$ from the 1000 independent runs of {\tt HODmin} for the {\tt c000\_ph000}+5pAHOD interpolator against the {\tt c000\_ph000}+5pAHOD-set-A mock data vector ($r_\mathrm{min2}=5$, $\mu_\mathrm{max2}=0.9$, monopole+quadrupole). The two histograms correspond to the outputs of {\tt HODmin}-S1 and {\tt HODmin}-S2. {\tt HODmin}-S2 consistently converges well within $\chi^2_\nu \ll 1$.}
    \label{fig:jackknife_val}
\end{figure}

Figure~\ref{fig:jackknife_val} shows the distribution of minimum $\chi^2_\nu$ values across 1000 independent {\tt HODmin} runs for this configuration, with the {\tt HODmin}-S1 and {\tt HODmin}-S2 distributions overlaid. The {\tt HODmin}-S2 refinement consistently drives the solution well into the $\chi^2_\nu \ll 1$ regime, as expected for a self-consistent comparison.

\begin{figure*}[t!]
    \centering
    \includegraphics[width=\textwidth]{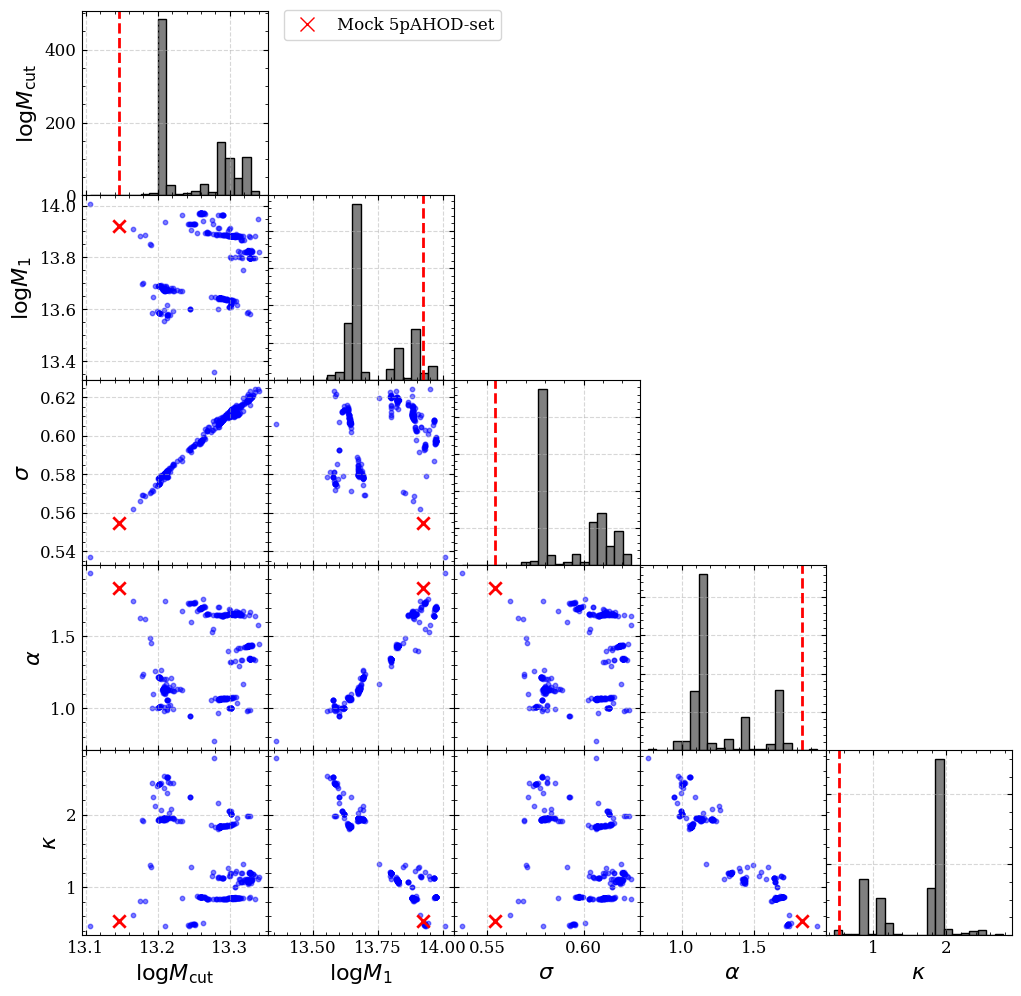}
    \caption{Scatter corner plot of the 1000 best-fit 5pIHOD-sets from the {\tt HODmin}-S2 outputs of the runs described in Figure~\ref{fig:jackknife_val}. Each of these 5pIHOD-sets corresponds to $\chi^2_\nu \ll 1$ relative to the same mock data vector. Thus, this distribution exposes a nontrivial degeneracy in 5pIHOD-space, which suggests an analogous degeneracy exists in 5pAHOD-space as well.}
    \label{fig:corner_val}
\end{figure*}

\begin{figure}[t!]
    \centering
    \includegraphics[width=\columnwidth]{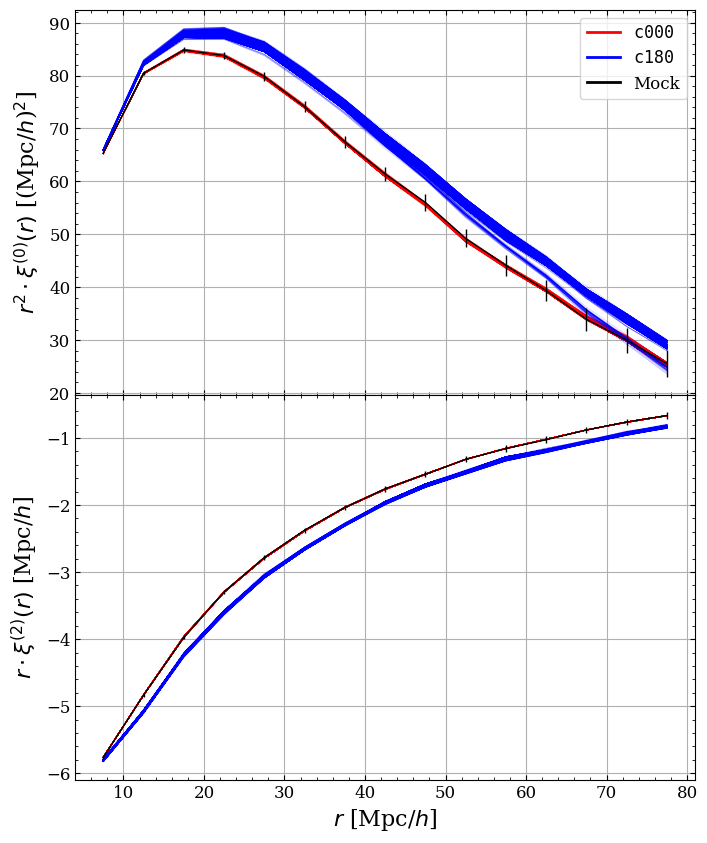}
    \caption{Two sets of 1000 2PCF multipoles corresponding to the {\tt HODmin}-S2 best-fit 5pIHOD-sets in comparison to the {\tt c000\_ph000}+5pAHOD-set-A mock data vector (black): {\tt c000\_ph000}+5pIHOD (red; corresponds to Figures~\ref{fig:jackknife_val}--\ref{fig:corner_val}, along with the mock data vector) and {\tt c180\_ph000}+5pIHOD (blue). Despite the wide spread in best-fit 5pIHOD-space (Figure~\ref{fig:corner_val}), the best-fit {\tt c000\_ph000}+5pIHOD multipoles appear nearly indistinguishable and correspond to $\chi^2_\nu \ll 1$. The {\tt c180\_ph000}+5pIHOD multipoles correspond to $\chi^2_\nu \sim 3$, and one can see that there exists an inherent difference in their shape compared to the mock data vector which cannot be overcome by the flexibility of the 5pIHOD model. 
    }
    \label{fig:ximock1000c000}
\end{figure}

The corresponding 1000 best-fit 5pIHOD-sets span a wide region of 5pIHOD space (Figure~\ref{fig:corner_val}). This shows that even in the self-consistent case, the profiled $\chi^2$ surface contains substantial degeneracies among the HOD parameters. 
It is not clear that any method of post-profile marginalization over these 1000 5pIHOD-sets would yield a meaningful \enquote{summary} or \enquote{final} 5pIHOD-set, since the marginalization result quite plausibly could fall outside the nontrivial degeneracy shape and yield a poor fit to data.
If we cannot extract a summary 5pIHOD-set, then we certainly cannot extrapolate a meaningful summary 5pAHOD-set, which may be of interest in non-{\tt c000} cosmologies. This is evident from examining the $\mathrm{log}M_\mathrm{cut}-\sigma$ panel, where even in this self-consistent case, the true 5pAHOD-set of the mock data vector falls on the outskirts of the converged 5pIHOD-sets. These effects, which can also be observed in the remaining panels, result from some combination of discreteness due to finite volume and shot noise, interpolation errors, and the nontrivial nature of the 5pHOD model itself. Figure~\ref{fig:ximock1000c000} confirms that this wide spread in parameter space nevertheless corresponds to 2PCF multipoles that are nearly indistinguishable from the mock data vector within the error bars. Thus, the minimum $\chi^2$ is a useful measure of goodness of fit, but the best-fit HOD parameters themselves cannot be interpreted as uniquely determined.

For comparison, Figure~\ref{fig:ximock1000c000} also shows the same multipole plot for the {\tt c180\_ph000}+5pIHOD interpolator fit to the same {\tt c000\_ph000}+5pAHOD-set-A mock. In this case the best-fit value is $\chi^2_\nu\sim 3$. The best-fit multipoles still reproduce the data at the order-of-magnitude level, but the systematic offset visible in the figure produces the elevated $\chi^2_\nu$. This plot shows an example of a such a strong difference in underlying cosmology as to be irrecoverable by the flexibility of the 5pIHOD model.
%% histo of final redchisqs for 1000 of c000 v c000

%% 5pAHOD scatter corner v c000

%% xi02 v c000 v c180

\subsection{Constraints across the cosmology suite}
\label{sec:results:suite}

The cases shown in Section~\ref{sec:results:baseline} are sample cases out of the full 81-cosmology suite analysis. We now consider results from the full suite. All numerical values of $\chi^2_\nu$ for each cosmology herein are given in Table~\ref{tab:HODmin_fin}. Unless otherwise stated (\ref{sec:results:ph002}, \ref{sec:results:HOD}), each comparison uses the {\tt c000\_ph000}+5pAHOD-set-A mock data vector as the target and reports the minimum $\chi^2_\nu$ obtained after profiling over the 5pIHOD parameters. We first examine how the floor constraint depends on analysis choices before comparing it to the ceiling case.

\subsubsection{Scale-cut}

\begin{figure*}[t!]
    \includegraphics[width=\textwidth]{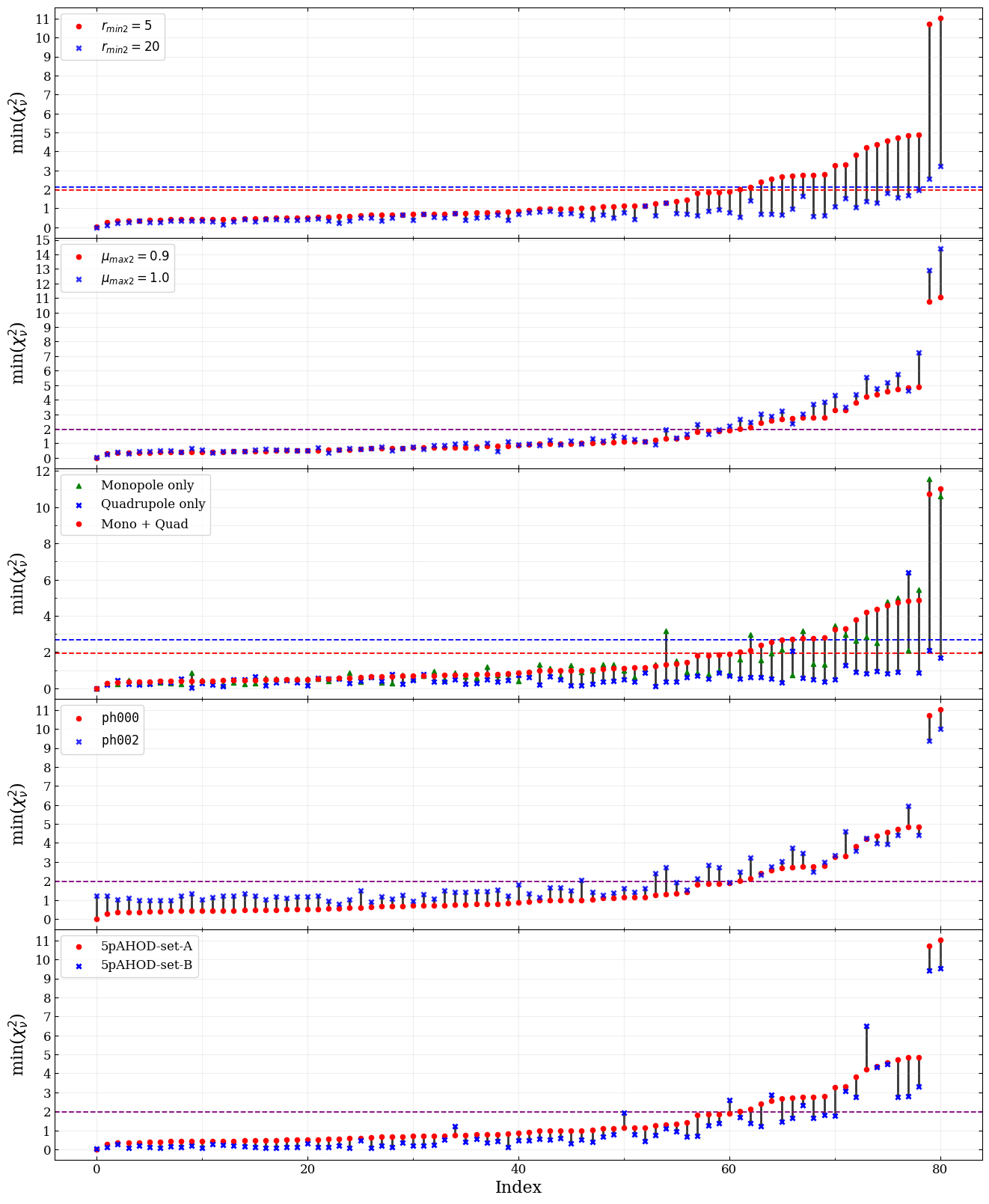}
    \centering
    \caption{
    The effect on suite constraints from varying the available inputs one at a time from the default test configuration of $r_\mathrm{min2}=5$, $\mu_\mathrm{max2}=0.9$, monopole+quadrupole, and a mock data vector produced from {\tt c000\_ph000}+5pAHOD-set-A (red). The first panel shows the effect of varying to $r_\mathrm{min2}=20$ (blue). The second panel shows the effect of varying to $\mu_\mathrm{max2}=1.0$ (blue). The third panel shows the effect of considering monopole only (green) and quadrupole only (blue). The fourth panel shows the effect of varying to mock phase {\tt ph002} (blue). The fifth panel shows the effect of varying to mock 5pAHOD-set-B (blue). The dashed lines on each subplot show the $\chi^2_\nu$ value corresponding to $3\sigma$ for the respective color under the assumption of Gaussian uncertainties. Indexing in each case is according to the default configuration, specifically by increasing minimum $\chi^2_\nu$. Thus, each index corresponds to the same cosmology across each of the five panels. The numerical values of $\chi^2_\nu$ for each cosmology are given in Table~\ref{tab:HODmin_fin}.
    }
    \label{fig:TCfin_linsuite_TOT}
\end{figure*}

Figure~\ref{fig:TCfin_linsuite_TOT} (first panel) shows the effect of varying the minimum separation included in the fit. Including the range $r\in[5,20]$ Mpc/$h$ generally strengthens the floor constraint, especially for cosmologies that are already in moderate tension when only $r\in[20,80]$ Mpc/$h$ is used. This occurs even though the $r_\mathrm{min2}=5$ case contains more data points and therefore has a larger number of degrees of freedom in the denominator of $\chi^2_\nu$. The result supports the general conclusion that small scales contain significant cosmological information, even in the conservative setting where the HOD parameters are profiled.

%% HODmin -- TOTAL

\subsubsection{$\mu$-cut}

Figure~\ref{fig:TCfin_linsuite_TOT} (second panel) shows the effect of varying the angular cut $\mu_\mathrm{max2}$. Increasing $\mu_\mathrm{max2}$ from 0.9 to 1.0 includes the most line-of-sight-oriented pairs, $\mu\in[0.9,1.0]$. This additional information tends to tighten the floor constraint modestly, but the effect is smaller than that obtained by lowering $r_\mathrm{min2}$ and is not uniform across the cosmology suite. The tightening is most visible for cosmologies that are already in moderate tension when only $\mu<0.9$ pairs are included. Although the difference in constraints is milder between these two cases, our results suggest it may be safer to omit the extreme angle galaxy pairs characterized by $\mu\in[0.9,1.0]$ in both modeling and data analysis, and therefore avoid line-of-sight systematic effects.

\subsubsection{Multipole choice}

Figure~\ref{fig:TCfin_linsuite_TOT} (third panel) shows the effect of fitting the monopole, the quadrupole, or both multipoles simultaneously. Using both multipoles generally gives the strongest floor constraint, indicating that it is more difficult to find a single 5pIHOD-set that fits both the monopole and quadrupole than to fit either multipole alone. In a few cases, one of the single-multipole fits appears to give a stronger constraint than the combined fit. These cases should be interpreted with care: the three choices correspond to different data vectors and different numbers of degrees of freedom, so $\chi^2_\nu$ is not a perfectly uniform comparison statistic across them. It is worth noting that the monopole appears to contain substantially more unique cosmological information than the quadrupole in the majority of our cosmologies. Nevertheless, including the quadrupole often strengthens monopole-only constraints considerably.

\subsubsection{Mock phase}
\label{sec:results:ph002}

Figure~\ref{fig:TCfin_linsuite_TOT} (fourth panel) shows the effect of changing the phase of the mock data vector from {\tt ph000} to {\tt ph002}, while keeping the fiducial cosmology and 5pAHOD-set fixed otherwise. The floor constraint generally tightens, except for cosmologies that were already strongly excluded. This indicates that phase variation changes the target multipoles in a way that cannot always be absorbed by varying the HOD parameters within the {\tt ph000}-trained 5pIHOD model. In other words, changing the mock phase can move the data vector away from the region of multipole space reachable by the interpolator trained on {\tt ph000}.

It is worth noting that this test only considers a single non-{\tt ph000} phase. We cannot extrapolate these conclusions to other phases in the {\tt \textsc{AbacusSummit}} suite, let alone general or atypical phases, without explicitly testing them. Moreover, this test simply shows that mock phase has a non-negligible effect on final constraints.

\subsubsection{Mock galaxy--halo connection}
\label{sec:results:HOD}

Figure~\ref{fig:TCfin_linsuite_TOT} (fifth panel) shows the effect of changing the galaxy--halo connection model used to generate the mock data vector, from 5pAHOD-set-A to 5pAHOD-set-B (Table~\ref{tab:5pAHOD-sets}). The resulting floor constraints shift substantially across the cosmology suite. This demonstrates that the conservative constraining power is not determined only by the cosmological separation from the fiducial model, but also by the location of the target galaxy sample within HOD space. Some HOD realizations produce 2PCF multipoles that are easier for other cosmologies to mimic after profiling, while others leave stronger residual cosmological differences. Note, again, that this is only a single alternative HOD realization; different HOD realizations may yield generally stronger or weaker constraints, but the effect is not negligible.

\subsubsection{Floor versus ceiling}
\label{sec:results:floor_ceiling}

\begin{figure*}[t!]
    \centering
    \includegraphics[width=\textwidth, height=0.41\textheight]{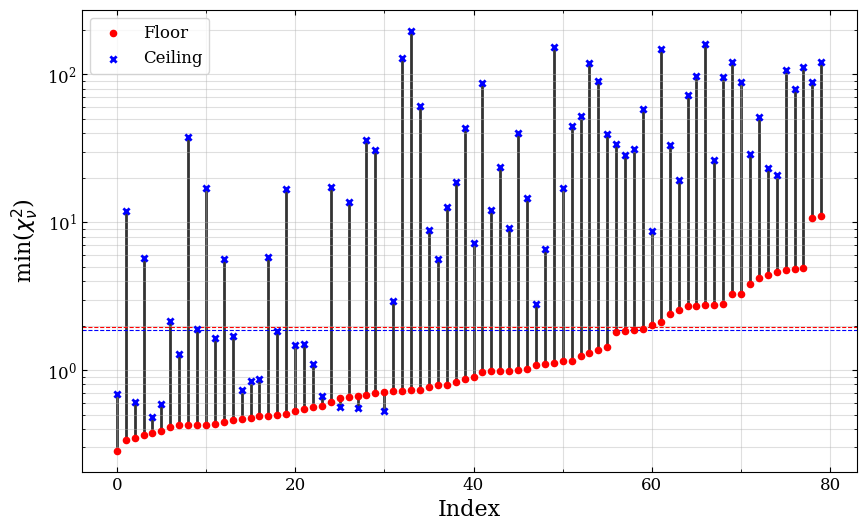}
    \caption{Floor versus ceiling constraints across the {\tt \textsc{AbacusSummit}} suite. 
    In the ceiling case (blue), computations are made directly from {\tt \textsc{AbacusHOD}} rather than through an interpolator, and no minimization is performed; 5pAHOD parameters are fixed to those of the mock data vector.
    The floor case (red), as well as the indexing of this plot, corresponds to the same default configuration shown in Figure~\ref{fig:TCfin_linsuite_TOT}.
    The floor--ceiling gap spans multiple orders of magnitude in $\chi^2_{\nu}$ and represents the range of constraints that could be produced given different 5pHOD priors within our otherwise-fixed methodology.
    Note that only 80 cosmologies are shown because the {\tt c000} case yielded such low $\chi^2_\nu$ values in both the ceiling (zero by tautology) and floor that these did not appear in the log-scaling. These cases, moreover, are omitted from our final computation of the percentage of cosmologies excluded at $3\sigma$ in the ceiling and floor cases: 20 of the 80 ($25\%$) floor cosmologies are excluded at $3\sigma$, whereas 65 of the 80 ($\sim81\%$) ceiling cosmologies are excluded at $3\sigma$. There are three cases visible where the floor exceeds the ceiling: this occurs because the model and mock cosmologies are so similar that their difference is less than the error due to interpolation and {\tt HODmin} convergence. The numerical values of $\chi^2_\nu$ for each cosmology are given in Table~\ref{tab:HODmin_fin}, and the floor results are plotted against the $w_0w_a$CDM parameter space in Figure~\ref{fig:TCfin_corner}.
    }
    \label{fig:TCfin_lin_ceil}
\end{figure*}

The 
primary
result of this paper is the comparison between the floor and ceiling constraints across the cosmology suite, shown in Figure~\ref{fig:TCfin_lin_ceil}. The ceiling case fixes the HOD parameters to those of the mock data vector and computes $\chi^2$ directly, with no minimization over HOD parameters. 
This represents the strongest constraint obtainable via the assumption that the exact HOD parameters are known. The floor case is the baseline {\tt HODmin} configuration, in which the HOD parameters are profiled over the broad prior bounds. 
This represents the weakest constraint obtainable, given our otherwise-fixed methodology described in Section~\ref{sec:Methods}, under the assumption that the data need only be fit by some allowed 5pHOD-set.

The gap between the two cases spans several orders of magnitude in $\chi^2_\nu$. With the mock data vector generated from the fiducial {\tt c000\_ph000} Planck $\Lambda$CDM cosmology, and under the assumption of Gaussian uncertainties quantified by our covariance matrix, approximately $81\%$ of the remaining 80 cosmologies are excluded at $3\sigma$ in the upper-limiting ceiling case, compared to only $25\%$ in the conservative floor case.
We must acknowledge the limitation that our 81-cosmology {\tt \textsc{AbacusSummit}} suite is not intended to completely represent the $w_0w_a$CDM cosmological parameter space, nor do we intend this discrete grid as a reliable substitute for continuous parameter constraints. Rather, we have considered the discrete cosmologies in order to avoid setting forth constraints that are highly sensitive to the underlying interpolation assumptions.

The cosmologies excluded in the ceiling case but not in the floor case are those for which the cosmological difference from the fiducial model can be largely absorbed by changing the HOD parameters. We find that many of these cosmologies which agree substantially well with data when allowed HOD freedom are excluded by multiple orders of magnitude in $\chi^2_\nu$ in the ceiling. Cosmologies excluded even in the floor case are sufficiently different from Planck $\Lambda$CDM that no allowed 5pIHOD-set can reproduce the mock data vector. On the contrary, when both the floor and ceiling are low, the non-{\tt c000} cosmology is so similar to {\tt c000} that applying the mock HOD model to the non-{\tt c000} cosmology yields a galaxy catalog largely similar to the mock catalog even before any optimization is required. Note that there are three cases visible where the floor constraint exceeds the ceiling constraint slightly. This occurs when the model and mock cosmologies are so similar that their difference is less than the error resulting from imperfect interpolation and {\tt HODmin} convergence.

The floor and ceiling cases bracket the constraining power obtainable under our assumed HOD model and parameter bounds. 
A small-scale clustering constraints analysis
with an informative, but not perfectly known, HOD prior, yet employing our otherwise fixed methodology,
should produce constraints that lie somewhere between the floor and ceiling. The large separation between these limiting cases demonstrates that galaxy--halo connection assumptions have a leading effect on the amount of cosmological information that can be robustly extracted from small-scale clustering.

%% fin lin suite -- ceil

\subsubsection{Cosmological parameters}

\begin{figure*}[t!]
    \centering
    \includegraphics[width=0.97\textwidth]{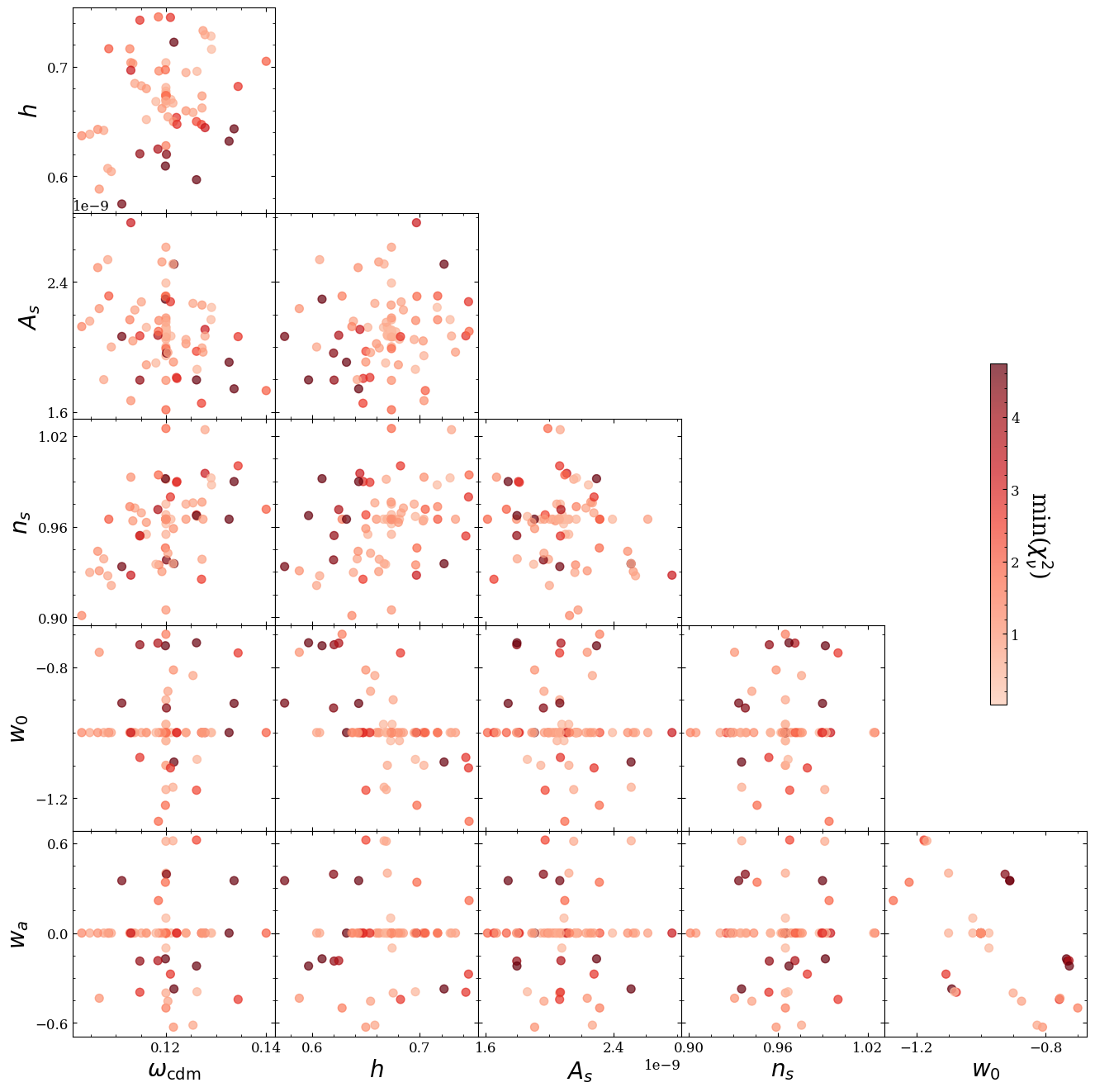}
    \caption{Baseline-configuration $\chi^2_\nu$ as a function of the cosmological parameters of the {\tt \textsc{AbacusSummit}}-based models. These are the same floor results presented linearly in Figure~\ref{fig:TCfin_lin_ceil}. The 81-cosmology sampling is too sparse for clear contours in general. With more data points, we might plausibly expect to uncover reliable contours in the $w_0-w_a$ and $h-\omega_\mathrm{cdm}$ projections, but it is not clear that this would help to produce coherent contours in the remaining projections unless we assumed an additional smoothing function that destroys cosmological information, which would contradict the purpose of this cosmology-information-maximizing floor test.}
    \label{fig:TCfin_corner}
\end{figure*}

Figure~\ref{fig:TCfin_corner} shows the baseline-configuration $\chi^2_\nu$ values, i.e. the floor results shown in Figure~\ref{fig:TCfin_lin_ceil}, projected against the cosmological parameters of the {\tt \textsc{AbacusSummit}} grid. This figure is intended as a diagnostic view of where the excluded and allowed cosmologies lie in the sampled parameter space. Although some structure appears to be plausibly discernible in the $w_0-w_a$ and $h-\omega_\mathrm{cdm}$ projections, the 81-cosmology grid is too sparse, and the floor values are scattered too inconsistently, to support reliable contours or a continuous interpolation of the constraint surface in general for all 2D projections without enforcing an additional smoothing assumption that destroys cosmological information. Since the primary goal of this floor test is to maximize cosmological information, we therefore restrict our conclusions to the discrete set of simulated cosmologies rather than attempting to infer continuous constraints on the cosmological parameters.

%% fin corner scatt

\subsection{Phase variation}
\label{sec:results:phase24}

\begin{figure*}[t!]
    \centering
    \includegraphics[width=0.8\textwidth]{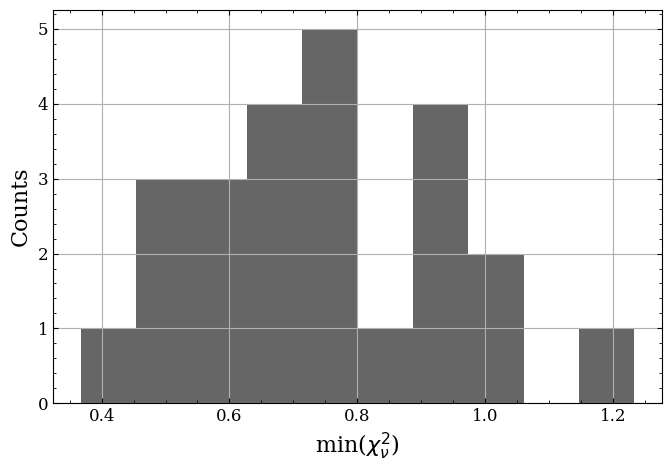}
    \caption{Distribution of minimum $\chi^2_\nu$ from {\tt HODmin} runs of the {\tt c000\_ph000}+5pAHOD interpolator against 24 mock data vectors from {\tt c000\_ph[001-024]}+5pAHOD-set-A. The 5pAHOD flexibility is generally insufficient to compensate phase differences within an otherwise-identical cosmology.}
    \label{fig:TCfin_24phases}
\end{figure*}

Finally, we test the sensitivity of the floor to phase variation within a single otherwise-fixed cosmology. We fit the {\tt c000\_ph000}+5pIHOD interpolator to 24 mock data vectors generated from {\tt c000\_ph[001-024]}+5pAHOD-set-A, as shown in Figure~\ref{fig:TCfin_24phases}. The resulting minimum $\chi^2_\nu$ values are nonzero and vary noticeably from phase to phase, especially compared to the near-zero minimum from the {\tt ph000} interpolator which is omitted in the Figure. This indicates that, at fixed cosmology and galaxy--halo connection, finite-volume phase variation changes the 2PCF multipoles in ways that cannot always be absorbed by varying the HOD parameters within the {\tt ph000}-trained 5pIHOD model.

%% phase min histo

%%%   %%%   %%%
%%%   %%%   %%%
%%%   %%%   %%%
% Section: Discussion
%%%   %%%   %%%
%%%   %%%   %%%
%%%   %%%   %%%
% Discussion
\section{Discussion and Conclusions}
\label{sec:discussion}

In this work we addressed the primary goal of comparing the strength of cosmological constraints, given some otherwise fixed methodology, when the non-cosmological contribution is absolutely minimized versus absolutely maximized. This allowed us to investigate how various galaxy--halo connection assumptions and methodological choices
affect cosmological constraints from small-scale galaxy clustering. We considered 2PCF monopoles and quadrupoles on scales ranging from $5$--$80$ Mpc/$h$, using mock LRG-like catalogs generated from the {\tt \textsc{AbacusSummit}} suite, and assuming the Planck $\Lambda$CDM cosmology at phase {\tt ph000} and an arbitrarily chosen galaxy--halo connection to be our fiducial model. We restricted the analysis to scales below $80$ Mpc/$h$ in order to focus on information beyond the standard BAO regime, and to scales above $5$ Mpc/$h$ to avoid the regime where the one-halo term dominates the cosmologies under consideration and results become exceedingly sensitive to the details of HOD modeling.

Our 
primary test
considered
how much cosmological constraining power remains when HOD parameters are treated as conservatively as possible, compared to the case where HOD parameters are known exactly. We addressed this by comparing two limiting cases: the \enquote{floor} and \enquote{ceiling}. In the floor case, we only impose broad bounds on our HOD parameters, along with modest filters on number density and satellite fraction, before profiling them with the {\tt HODmin} algorithm. A cosmology is excluded if no allowed HOD parametrization yields an acceptable fit to the mock data vector. The floor is deliberately conservative, insofar as no prior weight contributes to constraints, and we only seek whether any HOD parametrization makes the model viable. In the ceiling case, we assume the galaxy--halo connection is known exactly and compute the goodness of fit directly without any profiling or interpolation. This gives the strongest constraint obtainable under our modeling assumptions.

We examined the impact of several modeling and mock data vector choices on floor constraints. We found that substantial cosmological information exists in the small-scale regime of $r\in[5,20]$ Mpc/$h$, especially when cosmologies are already in at least moderate tension for $r\in[20,80]$ Mpc/$h$. This theory result agrees well with the BOSS LOWZ data result of \citet{lange_etal_2022}, who found that most of the multipole information in their {\tt \textsc{Aemulus}}-based clustering analysis fell within the range of [2,20] Mpc/$h$.
We found a smaller effect from including extreme-angle galaxy pairs in $\mu\in[0.9,1.0]$. The effect is nonzero, however, and it would seem to be safer to omit the contribution of extreme-angle pairs in both modeling and data analysis in order to avoid systematic effects which have been noted in numerous studies \citep{hawkins20032df,Guo_2012,Hahn_mumax,bianchi2017unbiased,mohammad2020completed}. We generally find that the monopole alone includes considerably more information than the quadrupole alone, but the inclusion of the quadrupole tends to tighten constraints considerably. We also find, as expected, that our results are sensitive to the phase and galaxy--halo connection by which the mock data vector is generated.

The difference between the limiting floor and ceiling cases is substantial. As shown in Figure~\ref{fig:TCfin_lin_ceil}, when we assume a Gaussian distribution of uncertainties in addition to our modeling assumptions enumerated in Section~\ref{sec:Methods:Procedure}, we find that 65 of the 80 non-fiducial cosmologies ($\sim81\%$) are excluded at $3\sigma$ in the ceiling case, compared to only 20 ($25\%$) in the floor case. Many of the cosmologies that agree considerably well with data in the floor case are excluded by multiple orders of magnitude in $\chi_\nu^2$ in the ceiling case. Thus, to the extent that the {\tt \textsc{AbacusSummit}} grid represents the $w_0w_a$CDM parameter space, the constraining power extractable from the substantial amount of cosmological information present in small-scale clustering depends strongly on the galaxy--halo connection assumptions set forth in modeling. Of course, our adherence to discretely sampled simulations does not allow us to infer continuous cosmological parameter constraints. However, this result is still noteworthy insofar as every {\tt \textsc{AbacusSummit}}-based small-scale clustering study in the literature deriving continuous cosmological parameter constraints is predicated on exactly the same discrete grid of simulations we have examined, whether directly or indirectly (e.g. \citealt{yuan_garrison_eisenstein_wechsler_2022,2024MNRAS.531..898P,2024PhRvD.109j3503V,2025arXiv251215962L}). Figure~\ref{fig:TCfin_corner} displays the floor results against the $w_0w_a$CDM parameter space, which we continue to discuss below. We also note that our floor--ceiling results agree qualitatively with \enquote{Figure 7} of \citet{2019MNRAS.484..989W} which, although predicated on an inherently distinct marginalization-based methodology that does not produce an HOD \enquote{floor}, displays how the loosening of assumed information leads to substantially weaker final constraints from small-scale clustering. Therein \citet{2019MNRAS.484..989W} finds a difference of about an order of magnitude in the fractional uncertainty on $\sigma_8$, whereas in our discrete-grid analysis we have found several instances of two to three orders of magnitude difference in absolute goodness of fit between floor and ceiling.

We found that the HOD parameter values associated with the minimum $\chi^2$ should not be interpreted as reliable estimates of the \enquote{true} or \enquote{final} HOD parameters in any meaningful sense. In the self-consistent tests, many widely-separated HOD parameter sets produce nearly indistinguishable 2PCF multipoles for considerably low $\chi^2$ values, as shown in Figures~\ref{fig:jackknife_val}--\ref{fig:ximock1000c000}. Consequently, profiling is useful for asking whether a cosmology can fit the data, but the corresponding best-fit parameters themselves should be interpreted with caution: a substantial and nontrivial degeneracy appears to exist within HOD space, such that it is not possible to infer an HOD model from its 2PCF multipoles alone, even when the cosmology is otherwise known completely. This degeneracy, which arises due to some combination of finite volume, shot noise, interpolation error, and the underlying 5pHOD model, seems to prevent any attempt at extracting summary HOD parameters.

It is worth reminding the reader that all $\chi^2$ computations in this work depend on the assumption that the covariance matrices derived from the Planck $\Lambda$CDM phases suffice for all interpolation and data vectors, whether or not they pertain to the Planck $\Lambda$CDM model. Although this approximation may suffice in low-$\chi^2$ cases, we have not validated the applicability in general, such that the precise numerical values in the high-$\chi^2$ cases are correspondingly affected. We do not expect that the floor--ceiling gap, which spans multiple orders of magnitude, will be significantly altered by modest adjustments to our covariance matrices. Moreover, 
in this particular test
we are not 
interested in precise numerical values, but rather in the difference that arises between floor and ceiling under otherwise identical modeling and covariance assumptions.

We chose to restrict our conclusions to the discrete set of {\tt \textsc{AbacusSummit}} cosmologies analyzed here. We attempted to interpolate the goodness-of-fit surface across the continuous cosmological parameter space in preparation, but found the result to be quite sensitive to essentially arbitrary and non-cosmological choices in the multidimensional interpolation scheme. It is not difficult to see why this is the case from Figure~\ref{fig:TCfin_corner}, where we show the baseline floor results. One might argue reasonably that smooth contours could be deduced in the $w_0-w_a$ and $h-\omega_\mathrm{cdm}$ projections. Clearly, however, smooth contours cannot be deduced in the majority of our remaining projections, not only due to the sparseness of the grid, but more importantly to the inconsistency of the floor values themselves. It would seem that smooth contours could only be extracted by positing an additional, non-cosmological smoothing assumption that destroys cosmological information. This would be antithetical to the primary goal of this floor test, which is to provide cosmology-information-maximizing constraints. We therefore report consistency tests only for the available cosmologies, rather than set forth continuous cosmological contours. 
Nevertheless, in future work we will investigate the prospects of performing a reliable interpolation over continuous cosmological parameter space, especially as we turn to a real DESI data analysis.
Given our cosmology-maximizing floor results in Figures~\ref{fig:TCfin_lin_ceil} and \ref{fig:TCfin_corner}, we are motivated in particular to investigate the extent to which the many data-based claims of continuous and strong cosmological contours in the simulation-based small-scale clustering literature arise from additional non-cosmological assumptions.

Our primary conclusion is that HOD assumptions are a leading contributor in the extraction of cosmological information from small-scale galaxy clustering. If the galaxy--halo connection is assumed to be known, then small-scale 2PCF multipoles can strongly discriminate between cosmologies in the {\tt \textsc{AbacusSummit}} grid. If, instead, HOD parameters are allowed to vary freely within broad initial bounds, much of this constraining power is absorbed by the galaxy--halo connection. Any small-scale clustering analysis with informative but imperfect HOD priors built on our otherwise-fixed methodology should lie between these two limits.

%%%   %%%   %%%
%%%   %%%   %%%
%%%   %%%   %%%
% Section: Acknowledgments

%\begin{minipage}{\columnwidth}
\section*{Acknowledgments}

This research used resources of the National Energy Research Scientific Computing Center (NERSC), a Department of Energy User Facility using NERSC award HEP-ERCAP 36297 (desi-2026).

NM gratefully acknowledges support from the Department of Energy Office of Science grant DE-SC0011840, and would like to thank Bharat Ratra, Glenn Horton-Smith,  Giorgi Khomeriki, and Davit Modebadze for helpful discussions. 

LS gratefully acknowledges support from the NASA ROSES grant 12-EUCLID12-0004. 

ZB and LS gratefully acknowledge funding from the NASA Grant $\#$80NSSC24M0021, \enquote{Project Infrastructure for the Roman Galaxy Redshift Survey}.

The authors would like to thank the anonymous referees for helpful feedback that greatly improved this manuscript.

%\end{minipage}

%%%   %%%   %%%
%%%   %%%   %%%
%%%   %%%   %%%
% Bibliography
\bibliographystyle{aasjournal}
\bibliography{MAIN}   % give same name for bib file as the main tex file... req for arxiv

%%%   %%%   %%%
%%%   %%%   %%%
%%%   %%%   %%%
% Appendix
%%%   %%%   %%%
%%%   %%%   %%%
%%%   %%%   %%%
% Appendix -- setup commands
\appendix
\renewcommand{\thesection}{\Alph{section}}
\renewcommand{\thesubsection}{\thesection.\arabic{subsection}}
\renewcommand{\thesubsubsection}{\thesubsection.\arabic{subsubsection}}
% -------- SECTION --------
\let\oldsection\section
\renewcommand{\section}[1]{%
  \refstepcounter{section}%
  \setcounter{subsection}{0}%
  \setcounter{subsubsection}{0}%
  \begin{center}
    {\normalfont\bfseries\normalsize \thesection.\ #1}
  \end{center}%
  \addcontentsline{toc}{section}{\thesection.\ #1}%
}
% -------- SUBSECTION --------
\let\oldsubsection\subsection
\renewcommand{\subsection}[1]{%
  \refstepcounter{subsection}%
  \setcounter{subsubsection}{0}%
  \begin{center}
  {\normalfont\bfseries\normalsize \thesubsection.\ #1}%
  \end{center}
  \addcontentsline{toc}{subsection}{\thesubsection.\ #1}%
}
% -------- SUBSUBSECTION --------
\let\oldsubsubsection\subsubsection
\renewcommand{\subsubsection}[1]{%
  \refstepcounter{subsubsection}%
  \begin{center}
  {\normalfont\bfseries\normalsize \thesubsubsection.\ #1}%
  \end{center}
  \addcontentsline{toc}{subsubsection}{\thesubsubsection.\ #1}%
}

%%%   %%%   %%%
%%%   %%%   %%%
%%%   %%%   %%%
% Appendix -- Results
\section{Numerical Results}
\label{app:results}

In this Appendix we provide the numerical values corresponding to our main results shown in Figures~\ref{fig:TCfin_linsuite_TOT}--\ref{fig:TCfin_lin_ceil}. These results are compiled in Table~\ref{tab:HODmin_fin}, and to avoid an exceedingly long table caption, we describe the contents of the Table here.

Each entry in Columns A-H is the $\chi^2_\nu$ value that appears in one of the plots in Figures~\ref{fig:TCfin_linsuite_TOT}--\ref{fig:TCfin_lin_ceil}. Each column corresponds to one test configuration for each cosmology in the suite, where the {\tt \textsc{AbacusSummit}} cosmology identifiers are listed in the first column labeled \enquote{ID} ({\tt ph000} implied for each). These cosmologies are listed at \url{https://abacussummit.readthedocs.io/en/latest/simulations.html}.

Column A corresponds to baseline configuration of $r_\mathrm{min2}=5$, $\mu_\mathrm{max2}=0.9$, monopole+quadrupole, and a mock data vector produced from {\tt c000\_ph000}+5pAHOD-set-A. 
This default configuration is plotted in each panel of Figure~\ref{fig:TCfin_linsuite_TOT}, as well as Figure~\ref{fig:TCfin_lin_ceil}.
Each of the next six columns considers a single variation about this default configuration.

Column B corresponds to $r_\mathrm{min2}=20$, and is shown in the first panel of Figure~\ref{fig:TCfin_linsuite_TOT}.

Column C corresponds to $\mu_\mathrm{max2}=1.0$, and is shown in the second panel of Figure~\ref{fig:TCfin_linsuite_TOT}.

Column D corresponds to monopole-only, and is shown in the third panel of Figure~\ref{fig:TCfin_linsuite_TOT}.

Column E corresponds to quadrupole-only, and is shown in the third panel of Figure~\ref{fig:TCfin_linsuite_TOT}.

Column F corresponds to {\tt ph002}, and is shown in the fourth panel of Figure~\ref{fig:TCfin_linsuite_TOT}.

Column G corresponds to 5pAHOD-set-B, and is shown in the fifth panel of Figure~\ref{fig:TCfin_linsuite_TOT}.

Column H corresponds to the \enquote{ceiling}, and is shown in Figure~\ref{fig:TCfin_lin_ceil}.

The number given in parentheses next to each column identifier is $\nu$, the number of degrees of freedom for that configuration. Hence, multiplying an entry from a column by its $\nu$-value yields the original $\chi^2$ extracted directly from {\tt HODmin}. 

%%%   %%%   %%%
%%%   %%%   %%%
%%%   %%%   %%%
% HODmin Final Results Table
%\captionsetup{width=\textwidth}
\begin{longtable}{ccccccccc}
\caption{\\
This Table is referenced for two purposes: first, the Column labeled \enquote{ID} provides the list of cosmology IDs that identify all of the {\tt \textsc{AbacusSummit}} cosmologies used in our 81-cosmology analysis; second, we show numerical
results from our main tests in Figures~\ref{fig:TCfin_linsuite_TOT}--\ref{fig:TCfin_lin_ceil}. To avoid an exceedingly long table caption, see the main text of Appendix~\ref{app:results} for a description to understand these results.
}
\label{tab:HODmin_fin}\\
\hline
ID & A (25) & B (19) & C (25) &  D (10) &  E (10) & F (25) & G (25) & H (30) \\
\hline
\endfirsthead
\hline
ID & A (25) & B (19) & C (25) & D (10) & E (10) & F (25) & G (25) & H (30) \\
\hline
\endhead
{\tt 000} & 0.0152846 & 0.0108746 &  0.042377 & 0.01185 & 0.014327 & 1.23385 & 0.0281741 & 0 \\
{\tt 001} &  0.96834 & 0.8403 & 0.87184 & 1.317454 & 0.198308 & 1.13114 & 0.555 & 87.5457 \\
{\tt 002} & 1.43468 &  0.6926 &  1.6651 & 0.908127 & 0.623482 & 1.5299 & 0.67412 &  39.668 \\
{\tt 003} & 0.339038 & 0.249587 &  0.41147 & 0.261169 & 0.4415 & 1.0271 & 0.258954 & 11.8756 \\
{\tt 004} & 1.14648 & 0.439664 & 1.26412 & 0.594572 & 0.380369 & 1.42088 & 0.803532 & 16.9643 \\
{\tt 100} & 0.487536 & 0.422174 &  0.56762 & 0.540759 & 0.334941 & 1.19933 & 0.081828 & 0.87015 \\
{\tt 101} & 0.56652 &  0.2376 &  0.5557 & 0.589814 & 0.523341 & 0.80134 & 0.176999 & 1.094 \\
{\tt 102} & 1.02276 & 0.417679 & 1.31069 & 0.992679 & 0.262871 & 1.43487 & 0.382343 & 14.5372 \\
{\tt 104} &  0.42926 &  0.16786 & 0.474 & 0.218316 & 0.134218 & 1.23333 & 0.222056 & 1.63831 \\
{\tt 105} &  0.54675 & 0.368876 & 0.362084 & 0.401195 & 0.550787 & 0.947632 & 0.099424 & 1.51161 \\
{\tt 106} &  0.49200 & 0.405463 &  0.5695 & 0.494388 & 0.463944 &  1.0834 & 0.128086 & 5.79547 \\
{\tt 107} &  0.44420 & 0.325127 &  0.44037 & 0.321353 & 0.493055 & 1.20715 & 0.194868 & 5.633 \\
{\tt 109} & 0.504716 &  0.42439 &  0.51711 & 0.547155 & 0.166955 & 1.18882 &  0.32836 & 16.7766 \\
{\tt 110} & 0.645072 & 0.510608 & 0.66946 & 0.661064 & 0.606728 & 0.90262 & 0.0848812 & 0.559063 \\
{\tt 111} &  0.67009 & 0.518135 &  0.53331 & 0.303225 & 0.780193 & 1.0718 & 0.115368 & 0.5532 \\
{\tt 112} & 0.46158 & 0.410599 &  0.435 & 0.244111 & 0.477673 & 1.32702 & 0.147938 & 1.70293 \\
{\tt 113} & 0.499364 & 0.387309 & 0.518296 & 0.517672 & 0.316875 & 1.19351 & 0.1212 & 1.83977 \\
{\tt 114} & 0.712132 &  0.69055 & 0.597772 & 0.707842 & 0.775142 & 1.30469 & 0.185727 &  0.52528 \\
{\tt 115} &  0.47283 & 0.410749 &  0.61293 & 0.568778 & 0.160109 & 1.0088 &  0.096044 &  0.83952 \\
{\tt 116} & 0.42724 & 0.319202 & 0.379885 & 0.428506 & 0.211569 & 1.13733 & 0.2612 & 16.9771 \\
{\tt 117} &  0.42285 &  0.36529 &  0.42677 & 0.248541 & 0.529174 & 1.22494 & 0.133092 & 1.2761 \\
{\tt 119} &  0.46595 &  0.32942 & 0.545248 & 0.268852 & 0.655995 & 1.22678 & 0.101028 &  0.73683 \\
{\tt 120} & 0.346726 & 0.260626 & 0.293748 & 0.449094 & 0.250472 & 1.08741 &  0.071425 &  0.60946 \\
{\tt 121} & 0.4257 &  0.34825 & 0.538404 & 0.434158 & 0.303833 & 1.03898 & 0.07629 & 1.8915 \\
{\tt 122} &  0.53061 & 0.471539 & 0.739592 & 0.538128 & 0.572778 & 1.22136 & 0.135016 & 1.47815 \\
{\tt 123} & 0.377452 & 0.266258 & 0.45222 & 0.28659 & 0.228263 & 0.97478 & 0.126772 & 0.48227 \\
{\tt 124} & 0.389476 & 0.288926 & 0.50514 & 0.30636 & 0.36538 & 0.97366 &  0.093317 & 0.585607 \\
{\tt 125} & 0.285664 & 0.127617 & 0.246769 & 0.299736 & 0.182592 & 1.22018 &  0.10449 &  0.69372 \\
{\tt 126} &  0.575 &  0.36674 & 0.669752 & 0.868967 & 0.293072 & 1.02785 &  0.086747 &  0.66813 \\
{\tt 130} & 1.90233 &  0.79636 & 2.23316 & 0.827322 & 0.697735 & 1.91525 & 2.61871 &  57.772 \\
{\tt 131} & 2.11111 & 1.40926 & 2.45596 & 2.985717 & 0.635038 & 3.24576 & 1.39077 & 149.156 \\
{\tt 132} & 1.00037 &  0.62538 &  0.98802 & 0.898647 & 0.172005 & 2.03328 & 0.52668 &  40.285 \\
{\tt 133} &  0.86386 &  0.70299 & 0.93958 & 0.427644 & 0.755047 & 1.79704 &  0.47082 & 43.1 \\
{\tt 134} & 4.8644 & 1.95954 &  7.2745 & 5.436751 & 0.87677 &  4.4004 & 3.30496 & 111.315 \\
{\tt 135} & 1.79555 &  0.63696 & 2.32955 & 0.862285 & 0.695748 & 2.11138 &  0.71276 & 33.857 \\
{\tt 136} & 0.98864 &  0.70813 &  0.93054 & 0.647169 & 0.504915 & 1.63644 &  0.57294 & 23.5343 \\
{\tt 137} & 2.55176 &  0.71800 & 2.85619 & 1.929087 & 0.522019 & 2.75578 & 2.86562 & 19.3141 \\
{\tt 138} &  0.79873 &  0.65385 &  0.4632 & 0.733172 & 0.35096 & 1.53118 &  0.41660 & 12.6513 \\
{\tt 139} & 1.35744 &  0.74476 & 1.3815 & 1.541335 & 0.367829 & 1.94644 &  0.96288 &  89.9304 \\
{\tt 140} &  2.7696 &  0.59113 & 3.69827 & 1.345361 & 0.503541 & 2.49422 & 1.66662 & 26.3232 \\
{\tt 141} &  0.79402 & 0.546974 & 1.03466 & 1.173943 & 0.479597 & 1.46825 & 0.340353 & 5.6667 \\
{\tt 142} & 1.12237 &  0.79539 & 1.44895 & 0.987631 & 0.483415 & 1.61412 & 1.9307 & 151.777 \\
{\tt 143} & 3.2712 & 1.08027 & 4.32296 & 3.460567 & 0.498689 & 3.3378 & 1.75692 & 120.444 \\
{\tt 144} & 3.29244 & 1.51771 & 3.50746 & 2.955758 & 1.265275 &  4.615 & 3.06086 &  88.026 \\
{\tt 145} & 1.2545 &  0.62916 & 0.926472 & 1.307979 & 0.133857 & 2.39304 & 0.76332 &  52.475 \\
{\tt 146} & 0.42404 & 0.359674 &  0.64042 & 0.844002 & 0.041831 & 1.32318 & 0.207196 &  37.804 \\
{\tt 147} & 1.09713 & 0.488469 & 1.52448 & 1.323668 & 0.391327 & 1.39329 &  0.7892 & 6.58143 \\
{\tt 148} &  4.3741 &  1.3002 & 4.80196 & 2.521568 & 0.956748 & 3.98838 & 4.34564 & 23.2869 \\
{\tt 149} &  0.65667 & 0.33195 & 0.791184 & 0.343765 & 0.41381 & 1.17967 & 0.186999 & 13.7709 \\
{\tt 150} &  4.8291 & 1.67931 & 4.61684 & 2.098451 & 6.390027 & 5.94416 & 2.78436 & 79.213 \\
{\tt 151} & 0.73766 & 0.403201 & 1.04852 & 0.666006 & 0.255083 & 1.42884 & 0.375034 & 60.631 \\
{\tt 152} &  4.2213 & 1.35507 &  5.5362 & 2.828345 & 0.814708 & 4.23736 &  6.497 & 51.309 \\
{\tt 153} &  4.741 & 1.58954 & 5.74308 & 5.003772 & 0.887661 & 4.42336 & 2.77009 & 106.462 \\
{\tt 154} & 2.74837 & 1.64722 & 3.03254 & 3.187156 & 0.590012 & 3.44794 & 2.3168 & 160.837 \\
{\tt 155} & 3.8039 & 1.06586 & 4.3726 & 2.623522 & 0.917105 & 3.58228 & 2.74759 & 28.9585 \\
{\tt 156} & 2.38979 &  0.70928 & 3.04251 & 1.57481 & 0.599054 & 2.30902 & 1.23453 & 33.3328 \\
{\tt 157} &  0.60783 & 0.523741 & 0.618496 & 0.382956 & 0.41169 & 1.5111 &  0.485 & 17.3704 \\
{\tt 158} & 0.41546 & 0.346887 &  0.49925 & 0.272721 & 0.333277 & 0.99428 & 0.148066 & 2.1376 \\
{\tt 159} & 11.0486 & 3.22752 & 14.4311 & 10.597813 & 1.683407 & 10.0189 &  9.5317 & 121.457 \\
{\tt 160} & 1.86314 & 0.9224 & 1.9436 & 0.966056 & 0.848654 & 2.71268 & 1.37502 & 31.3019 \\
{\tt 161} & 0.902864 & 0.799095 & 0.95462 & 0.82473 & 0.606968 & 1.32394 & 0.467636 & 7.20563 \\
{\tt 162} & 0.71942 & 0.487445 &  0.8508 & 0.50956 & 0.384121 & 1.5143 & 0.507384 & 128.173 \\
{\tt 163} & 0.362659 & 0.347797 &  0.45369 & 0.338882 & 0.195906 & 1.0027 & 0.205326 & 5.7101 \\
{\tt 164} & 0.717684 &  0.54759 & 0.856964 & 0.953439 & 0.349637 & 1.07758 & 0.245846 & 2.93699 \\
{\tt 165} & 0.70186 & 0.398633 &  0.7585 & 0.537928 & 0.45336 & 0.95706 & 0.176845 & 30.6013 \\
{\tt 166} &  0.77320 & 0.488642 & 0.64892 & 0.56887 & 0.283462 & 1.46976 & 0.5548 &  8.806 \\
{\tt 167} & 2.0145 & 0.538384 & 2.6579 & 1.605799 & 0.517992 & 2.46391 & 1.68861 & 8.7758 \\
{\tt 168} &  0.6804 &  0.68117 & 0.68134 & 0.740885 & 0.249525 & 1.25345 & 0.340835 &  35.794 \\
{\tt 169} & 2.79131 &  0.62702 & 3.84975 & 1.331562 & 0.356697 & 2.97593 & 1.81641 &  96.229 \\
{\tt 170} & 0.734904 &  0.74116 &  0.94836 & 0.880175 & 0.470236 & 1.42464 & 1.20424 & 197.321 \\
{\tt 171} &  0.98848 & 0.852611 & 1.21196 & 1.101311 & 0.672255 & 1.65433 &  0.50178 & 12.1202 \\
{\tt 172} & 1.31376 & 1.29769 & 1.9674 & 3.194412 & 0.378372 &  2.7186 & 1.10596 & 118.448 \\
{\tt 173} & 0.992044 & 0.73668 & 1.16248 & 1.292444 & 0.168837 & 1.48431 & 0.301989 &  9.1344 \\
{\tt 174} & 1.14898 & 1.12263 & 1.11482 & 1.056407 & 0.845792 & 1.60288 &  0.41443 & 44.9053 \\
{\tt 175} & 10.7349 & 2.56497 & 12.9184 & 11.583109 & 2.109406 & 9.3588 & 9.41436 &  88.67 \\
{\tt 176} & 1.83207 & 0.848795 & 1.61996 & 0.759789 & 0.533098 & 2.82287 & 1.24771 & 28.396 \\
{\tt 177} & 2.69175 &  0.67088 & 3.23872 & 2.131157 & 0.321134 & 3.02241 & 1.45026 &  72.827 \\
{\tt 178} & 0.82562 & 0.391017 & 1.10955 & 0.823054 & 0.45927 & 1.20816 & 0.119216 & 18.7949 \\
{\tt 179} & 1.08438 & 0.668058 & 1.1934 & 1.33293 & 0.375662 & 1.24179 & 0.6509 & 2.81521 \\
{\tt 180} &  2.7292 &  0.9885 & 2.34202 & 0.744812 & 2.078212 & 3.74723 & 1.64081 &  96.924 \\
{\tt 181} &  4.577 &  1.8043 &  5.2084 & 4.776355 & 0.822119 & 3.9457 & 4.50828 & 20.9322 \\
\hline
\end{longtable}

%%%   %%%   %%%
%%%   %%%   %%%
%%%   %%%   %%%
% End Document
\end{document}